\documentclass[aps,prc,twocolumn,superscriptaddress,showpacs]{revtex4}

\usepackage{graphicx}
\begin{document}
\bibliographystyle{apsrev.bst}

\title{Identified hadron transverse momentum spectra\\ in Au+Au collisions 
at $\sqrt{s_{_{NN}}}$=62.4 GeV}
\author{B.B.Back}
\affiliation{Argonne National Laboratory, Argonne, IL 60439-4843}
\author{M.D.Baker}
\affiliation{Brookhaven National Laboratory, Upton, NY 11973-5000}
\author{M.Ballintijn}
\affiliation{Massachusetts Institute of Technology, Cambridge, MA 
02139-4307}
\author{D.S.Barton}
\affiliation{Brookhaven National Laboratory, Upton, NY 11973-5000}
\author{R.R.Betts}
\affiliation{University of Illinois at Chicago, Chicago, IL 60607-7059}
\author{A.A.Bickley}
\affiliation{University of Maryland, College Park, MD 20742}
\author{R.Bindel}
\affiliation{University of Maryland, College Park, MD 20742}
\author{W.Busza}
\affiliation{Massachusetts Institute of Technology, Cambridge, MA 
02139-4307}
\author{A.Carroll}
\affiliation{Brookhaven National Laboratory, Upton, NY 11973-5000}
\author{Z.Chai}
\affiliation{Brookhaven National Laboratory, Upton, NY 11973-5000}
\author{M.P.Decowski}
\affiliation{Massachusetts Institute of Technology, Cambridge, MA 
02139-4307}
\author{E.Garc\'{\i}a}
\affiliation{University of Illinois at Chicago, Chicago, IL 60607-7059}
\author{T.Gburek}
\affiliation{Institute of Nuclear Physics PAN, Krak\'{o}w, Poland}
\author{N.George}
\affiliation{Brookhaven National Laboratory, Upton, NY 11973-5000}
\author{K.Gulbrandsen}
\affiliation{Massachusetts Institute of Technology, Cambridge, MA 
02139-4307}
\author{C.Halliwell}
\affiliation{University of Illinois at Chicago, Chicago, IL 60607-7059}
\author{J.Hamblen}
\affiliation{University of Rochester, Rochester, NY 14627}
\author{M.Hauer}
\affiliation{Brookhaven National Laboratory, Upton, NY 11973-5000}
\author{C.Henderson}
\affiliation{Massachusetts Institute of Technology, Cambridge, MA 
02139-4307}
\author{D.J.Hofman}
\affiliation{University of Illinois at Chicago, Chicago, IL 60607-7059}
\author{R.S.Hollis}
\affiliation{University of Illinois at Chicago, Chicago, IL 60607-7059}
\author{R.Ho\l y\'{n}ski}
\affiliation{Institute of Nuclear Physics PAN, Krak\'{o}w, Poland}
\author{B.Holzman}
\affiliation{Brookhaven National Laboratory, Upton, NY 11973-5000}
\author{A.Iordanova}
\affiliation{University of Illinois at Chicago, Chicago, IL 60607-7059}
\author{E.Johnson}
\affiliation{University of Rochester, Rochester, NY 14627}
\author{J.L.Kane}
\affiliation{Massachusetts Institute of Technology, Cambridge, MA 
02139-4307}
\author{N.Khan}
\affiliation{University of Rochester, Rochester, NY 14627}
\author{P.Kulinich}
\affiliation{Massachusetts Institute of Technology, Cambridge, MA 
02139-4307}
\author{C.M.Kuo}
\affiliation{National Central University, Chung-Li, Taiwan}
\author{W.T.Lin}
\affiliation{National Central University, Chung-Li, Taiwan}
\author{S.Manly}
\affiliation{University of Rochester, Rochester, NY 14627}
\author{A.C.Mignerey}
\affiliation{University of Maryland, College Park, MD 20742}
\author{R.Nouicer}
\affiliation{Brookhaven National Laboratory, Upton, NY 11973-5000}
\affiliation{University of Illinois at Chicago, Chicago, IL 60607-7059}
\author{A.Olszewski}
\affiliation{Institute of Nuclear Physics PAN, Krak\'{o}w, Poland}
\author{R.Pak}
\affiliation{Brookhaven National Laboratory, Upton, NY 11973-5000}
\author{C.Reed}
\affiliation{Massachusetts Institute of Technology, Cambridge, MA 
02139-4307}
\author{C.Roland}
\affiliation{Massachusetts Institute of Technology, Cambridge, MA 
02139-4307}
\author{G.Roland}
\affiliation{Massachusetts Institute of Technology, Cambridge, MA 
02139-4307}
\author{J.Sagerer}
\affiliation{University of Illinois at Chicago, Chicago, IL 60607-7059}
\author{H.Seals}
\affiliation{Brookhaven National Laboratory, Upton, NY 11973-5000}
\author{I.Sedykh}
\affiliation{Brookhaven National Laboratory, Upton, NY 11973-5000}
\author{C.E.Smith}
\affiliation{University of Illinois at Chicago, Chicago, IL 60607-7059}
\author{M.A.Stankiewicz}
\affiliation{Brookhaven National Laboratory, Upton, NY 11973-5000}
\author{P.Steinberg}
\affiliation{Brookhaven National Laboratory, Upton, NY 11973-5000}
\author{G.S.F.Stephans}
\affiliation{Massachusetts Institute of Technology, Cambridge, MA 
02139-4307}
\author{A.Sukhanov}
\affiliation{Brookhaven National Laboratory, Upton, NY 11973-5000}
\author{M.B.Tonjes}
\affiliation{University of Maryland, College Park, MD 20742}
\author{A.Trzupek}
\affiliation{Institute of Nuclear Physics PAN, Krak\'{o}w, Poland}
\author{C.Vale}
\affiliation{Massachusetts Institute of Technology, Cambridge, MA 
02139-4307}
\author{G.J.van~Nieuwenhuizen}
\affiliation{Massachusetts Institute of Technology, Cambridge, MA 
02139-4307}
\author{S.S.Vaurynovich}
\affiliation{Massachusetts Institute of Technology, Cambridge, MA 
02139-4307}
\author{R.Verdier}
\affiliation{Massachusetts Institute of Technology, Cambridge, MA 
02139-4307}
\author{G.I.Veres}
\affiliation{Massachusetts Institute of Technology, Cambridge, MA 
02139-4307}
\author{E.Wenger}
\affiliation{Massachusetts Institute of Technology, Cambridge, MA 
02139-4307}
\author{F.L.H.Wolfs}
\affiliation{University of Rochester, Rochester, NY 14627}
\author{B.Wosiek}
\affiliation{Institute of Nuclear Physics PAN, Krak\'{o}w, Poland}
\author{K.Wo\'{z}niak}
\affiliation{Institute of Nuclear Physics PAN, Krak\'{o}w, Poland}
\author{B.Wys\l ouch}
\affiliation{Massachusetts Institute of Technology, Cambridge, MA 
02139-4307}

\collaboration{PHOBOS Collaboration}
\noaffiliation
\date{\today}
\begin{abstract}
Transverse momentum spectra of pions, kaons, protons and antiprotons
from Au+Au collisions at $\sqrt{s_{_{NN}}}$ = 62.4 GeV have been measured by 
the PHOBOS experiment at the Relativistic Heavy Ion Collider 
at Brookhaven National Laboratory. 
The identification of particles relies on three different methods: 
low momentum particles stopping in the first detector layers; 
the specific energy loss ($dE/dx$) in the silicon Spectrometer, 
and Time-of-Flight measurement.
These methods cover the transverse momentum ranges 
0.03--0.2, 0.2--1.0 and 0.5--3.0~GeV/c, respectively. 
Baryons are found to have substantially harder 
transverse momentum spectra than mesons. 
The $p_T$ region in which the proton to pion ratio reaches unity in  
central Au+Au collisions at $\sqrt{s_{_{NN}}}$ = 62.4 GeV
fits into a smooth trend as a function of collision energy. 
At low transverse momenta, the spectra exhibit a significant 
deviation from transverse mass scaling,
and when the observed particle yields at very low $p_T$ are compared to
extrapolations from higher $p_T$, 
no significant excess is found.
By comparing our results to Au+Au collisions at $\sqrt{s_{_{NN}}}$ = 200 GeV,
we conclude that the net proton yield at midrapidity is proportional to 
the number of participant nucleons in the collision.
\end{abstract}
\maketitle

\section{Introduction}

The yield of identified hadrons produced in collisions of gold nuclei
at an energy of $\sqrt{s_{_{NN}}}$ = 62.4 GeV has been measured with the 
PHOBOS detector at the Relativistic Heavy Ion Collider (RHIC) at Brookhaven 
National Laboratory. The data are presented as a function of transverse 
momentum, transverse mass, and centrality.

It was shown previously by PHOBOS  that, 
the 
inclusive charged hadron transverse momentum spectra in Au+Au collisions 
exhibit the same centrality dependence at center-of-mass 
energies of $\sqrt{s_{_{NN}}}$ = 200 GeV and 62.4 GeV \cite{phobos:spectra_AuAu_63}.
It is also known that in central 
Au+Au collisions at 200 GeV 
\cite{Adler:2003cb} and 130 GeV \cite{Adcox:2001mf,Adcox:2003nr},  
the proton and antiproton yields become comparable to the 
pion yields above $p_T\approx 2$~GeV/c.
The yield of high-$p_T$ particles has been measured to be suppressed 
with respect to the scaling with the number of binary nucleon-nucleon 
collisions \cite{Adcox:2001jp,Adcox:2002pe,Adler:2003au,
Adler:2002xw,Adams:2003kv,Arsene:2003yk,Back:2003qr}, but 
that suppression was found to be strongly species-dependent 
\cite{Adler:2003cb}.

The present study at 62.4 GeV aims to extend the energy range
for which the contributions of different particle species to the 
inclusive charged hadron spectra are known.
These results add to our knowledge of the energy dependence 
of baryon transport and baryon production in heavy-ion collisions, 
and of parton energy loss in the hot and dense medium \cite{Gyulassy:1990ye},
that is believed to be produced.

The results presented here allow the first examination of differences between 
the transverse dynamics of various particle species at 62.4 GeV, bridging a
gap between the top SPS energy ($\sqrt{s_{_{NN}}}$=17.2 GeV) and the 
higher RHIC energies (130 and 200 GeV). This is the first 
publication using the PHOBOS Time-of-Flight detector for particle 
identification, and using the PHOBOS silicon Spectrometer to obtain momentum spectra
of identified particles.


\section{The PHOBOS Detector}

\begin{figure*}[t]
\includegraphics[angle=0,width=138mm]{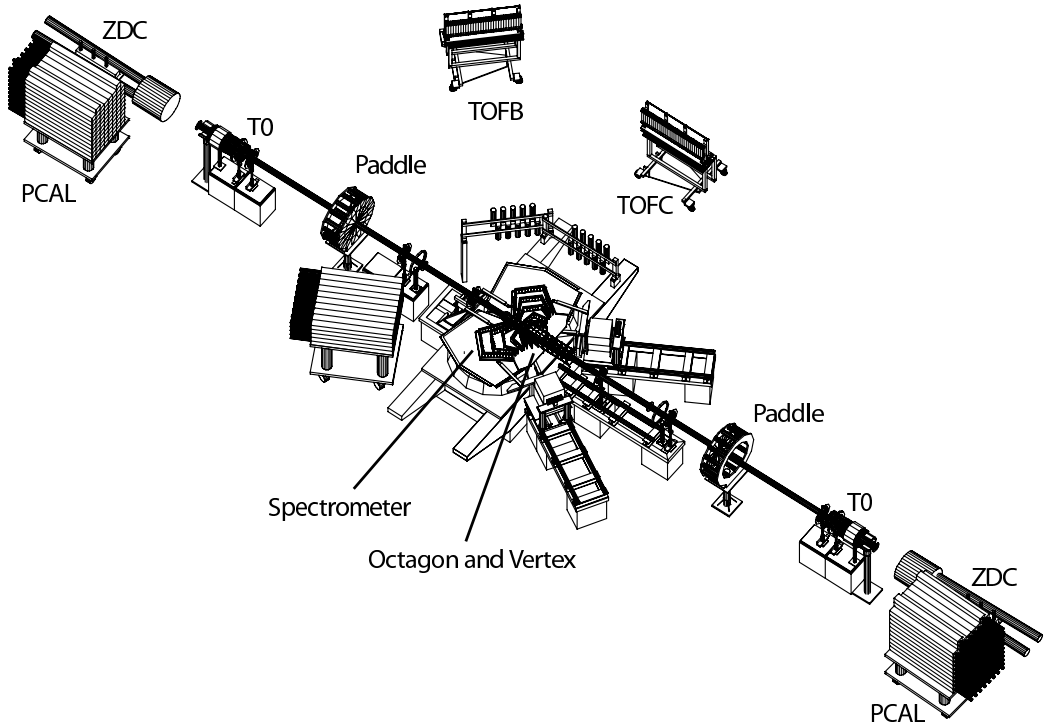}
\caption{\label{phobosdetector}
The layout of the PHOBOS detector system used in the Au+Au run at 62.4 GeV.
The  silicon Spectrometer arms are located in the center,
placed in a double dipole magnetic field. The two Time of Flight
scintillator walls (center-top of the figure) detect particles emitted 
near 45 and 90 degrees from the beam direction and are located at a 
distance to the collision point of
5.34 and 3.80 meters, respectively.}
\end{figure*}

 The PHOBOS detector \cite{phobos:detector} is designed to provide global 
 characterisation of heavy-ion collisions, with about 1\% of particles 
 analyzed in detail in the Spectrometer system.
 The layout of the PHOBOS detector system
 is shown in Fig.~\ref{phobosdetector}.
 Only the parts of the detector relevant to the present 
 analysis will be described.

\subsection{Event Trigger and Vertex-finding}

 The primary event trigger requires a coincidence between the `Paddle 
 Counters' \cite{phobos:paddles}, which are two sets of sixteen 
 scintillator detectors located at $z=\pm3.21$~m 
 (where $z$ is the distance from the nominal 
 collision point along the beam direction) 
 and spanning the pseudorapidity region $3.2 \le |\eta| \le 4.5$.

 The Zero-Degree Calorimeters (ZDCs) \cite{rhic:zdc}, positioned at 
 $\pm18$~m, measure spectator neutrons from the collision. With an 
 identical design for each of the four RHIC experiments, the ZDCs are 
 built from tungsten optical-fibre sandwiches. A requirement of a ZDC 
 coincidence can be added to the event trigger to enhance trigger purity 
 in high background situations.

 An online vertex is determined with a resolution of roughly $4$~cm using 
 the Time-Zero (T0) detectors, two sets of ten \v Cerenkov radiators 
 situated close to the beam pipe, at 
 $z = \pm5.2$~m. This vertex trigger enhances the 
 fraction of recorded events in the vertex region 
 in which the efficiency of the PHOBOS Spectrometer is maximal, 
 $-20 \le z \le 20$~cm.

Offline  vertex  reconstruction makes use of information from different 
sub-detectors. 
Two sets of double-layered silicon Vertex Detectors are located 
below and above the collision point. 
PHOBOS also has two Spectrometer 
arms in the horizontal plane used for tracking and momentum measurement 
of charged particles.
For events in the selected vertex region, the most 
accurate $z$ and $y$ (vertical) positions are obtained from the Vertex 
Detector, while the position along $x$ (horizontal, perpendicular to 
the beam) comes primarily from the Spectrometer. 
The final resolution of the vertex position along $z$ is found to be 
better than 300~$\mu$m.

\subsection{Particle Tracking and Identification Detectors}

 Particle tracking and identification in the PHOBOS experiment is performed using the 
 Spectrometer and the Time-of-Flight (TOF) system.

 Each arm of the Spectrometer consists of 137 silicon pixel 
 sensors arranged into layers, with an azimuthal angular coverage 
 of $\Delta \phi \approx 0.1$ radians. The silicon sensor technology used 
 in the PHOBOS detector is described in \cite{phobos:si}. The Spectrometer 
 sensors are designed to give precise hit position determination in the 
 $x-z$ plane.

 The first silicon layer is positioned within 10~cm from the interaction point, 
 allowing good rejection of tracks from displaced vertices. 
The thickness of the beam-pipe is 1 mm, and it is constructed from
beryllium to minimise multiple scattering and secondary particle 
production.

 The PHOBOS double-dipole magnetic field is designed such that the inner 
 six Spectrometer layers sit in a low-field region while the remaining 
 layers are in an approximately constant vertical magnetic field of 2~T.

 There are two TOF walls: Wall B, centered around a line at 45$^o$ to 
 the beam-line and located 5.4~m from the origin; and Wall C, also facing 
 the collision point, but centered around a line at 90$^o$ to the beam-line,
 at a distance of 3.9~m. Each wall consists of 120 vertical Bicron 
 BC-404 plastic scintillator rods which are 20~cm long with a 
 cross-section of 8$\times8$~$mm^2$. The scintillators have PMT read-out top 
 and bottom, providing vertical position information.
 The present analysis uses only data from TOF Wall B, due to low 
 statistics in Wall C for the short 62.4~GeV run.

 Particle timing information is obtained relative to an event `start-time' 
 determined from the T0 \v Cerenkov detectors. 
 Leading edge discriminators are used, therefore slew corrections of the 
 timing signals as a function of pulse height are necessary.
 The stability of the timing signals arriving at the TDCs over long cables
 is monitored and drifts caused by temperature changes are corrected for. 
 Channel-by-channel cable delay differences are corrected using TOF hits
 matched to tracks reconstructed in the Spectrometer with known momenta.
 A similar adjustment is accomplished for the T0 detectors using the 
 signals originating from the prompt (fastest) particles close to the beam 
 direction. The combined T0-TOF system was found to have a total 
 timing resolution of about 140~ps.
 

\section{Event Selection and Centrality}

The events selected for analysis were divided into three centrality classes, 
based on signals in the Paddle detectors.
The mean number of participating nucleons
($\langle {\rm N_{part}} \rangle$) and the mean number of binary
nucleon-nucleon collisions ($\langle {\rm N_{coll}} \rangle$)
were estimated for each centrality class using Monte-Carlo methods.
These $\rm{N_{part}}$ and $\rm{N_{coll}}$ values and their systematic
errors are listed in Table \ref{table1}.
Details of the estimation procedure and the selection of events 
can be found in Appendix \ref{appendix_evsel}.

\begin{table}
\begin{ruledtabular}
\begin{tabular}{lll}
Centrality &
$ \left<N_{part}\right>$ &  $ \left< N_{coll}\right>$ \\
\hline
30--50\% &  $88 \pm 9$    & $128  \pm 17$ \\
15--30\% & $175 \pm 9$    & $331  \pm 23$ \\
 0--15\% & $294 \pm 10$   & $684  \pm 23$ \\
\hline
 0--50\% & $176 \pm 9$    & $356  \pm 21$ \\
\end{tabular}
\end{ruledtabular}
\caption{
\label{table1}
Details of the centrality classes used in this analysis.
Bins are expressed in terms of percentage of the total inelastic Au+Au
cross-section at $\sqrt{s_{_{NN}}}$=62.4 GeV.}
\end{table}


\section{Track Reconstruction and Particle Identification}

 The present analysis uses essentially the same Spectrometer tracking 
 procedures as previously applied to obtain 
 non-identified charged hadron spectra results from 
 PHOBOS \cite{phobos:spectra_AuAu_200,phobos:spectra_dAu_200,phobos:spectra_dAu_forward,phobos:spectra_AuAu_63}. Details are given in Appendix~\ref{appendix_tracking}.


Three independent methods were used for particle
identification, over differing momentum ranges.

The charged particle yields at very low momentum were measured by searching 
for particle tracks that stop in the $5^{\rm th}$ spectrometer layer.
This technique allows identification of pions between 0.03 and 0.05 GeV/c, 
kaons between 0.09 and 0.13 GeV/c, and protons between 0.14 and 0.21 GeV/c
transverse momentum, close to mid-rapidity ($-$0.1$<$y$<$0.4).
The procedure is summarized in Appendix~\ref{appendix_sipid}, 
with further details presented in \cite{phobos:stopping}.

For particles at intermediate momentum, the velocity-dependent
specific energy loss ($dE/dx \propto 1/v^2$) in the silicon Spectrometer can be 
used to separate particles with the same momentum but different mass.
A truncated mean calculation is used, discarding the 30\% highest energy hits on each reconstructed Spectrometer track. 
This reduces the effect of the large Landau tails of the energy-loss distribution and improves sensitivity for particle identification.
Previous PHOBOS publications of antiparticle to particle ratios using 
$dE/dx$ identification \cite{phobos:pbarp_AuAu_130,phobos:pbarp_AuAu_200,phobos:pbarp_dAu_200,phobos:pbarp_pp_200} made cuts in $dE/dx$ versus total momentum $p$ to identify particles. 
In the present analysis, however, to extend the range for identification, particle yields were extracted as a function of momentum by fitting the $dE/dx$ distributions.
This is illustrated in Fig.~\ref{pidplot} and details are discussed in Appendix~\ref{appendix_sipid}.
This technique allows particle identification in the momentum ranges
$0.2 < p <0.9$~GeV/c for pions and kaons, and 
$0.3 < p < 1.4$~GeV/c for protons and antiprotons.

\begin{figure}
\includegraphics[width=87mm]{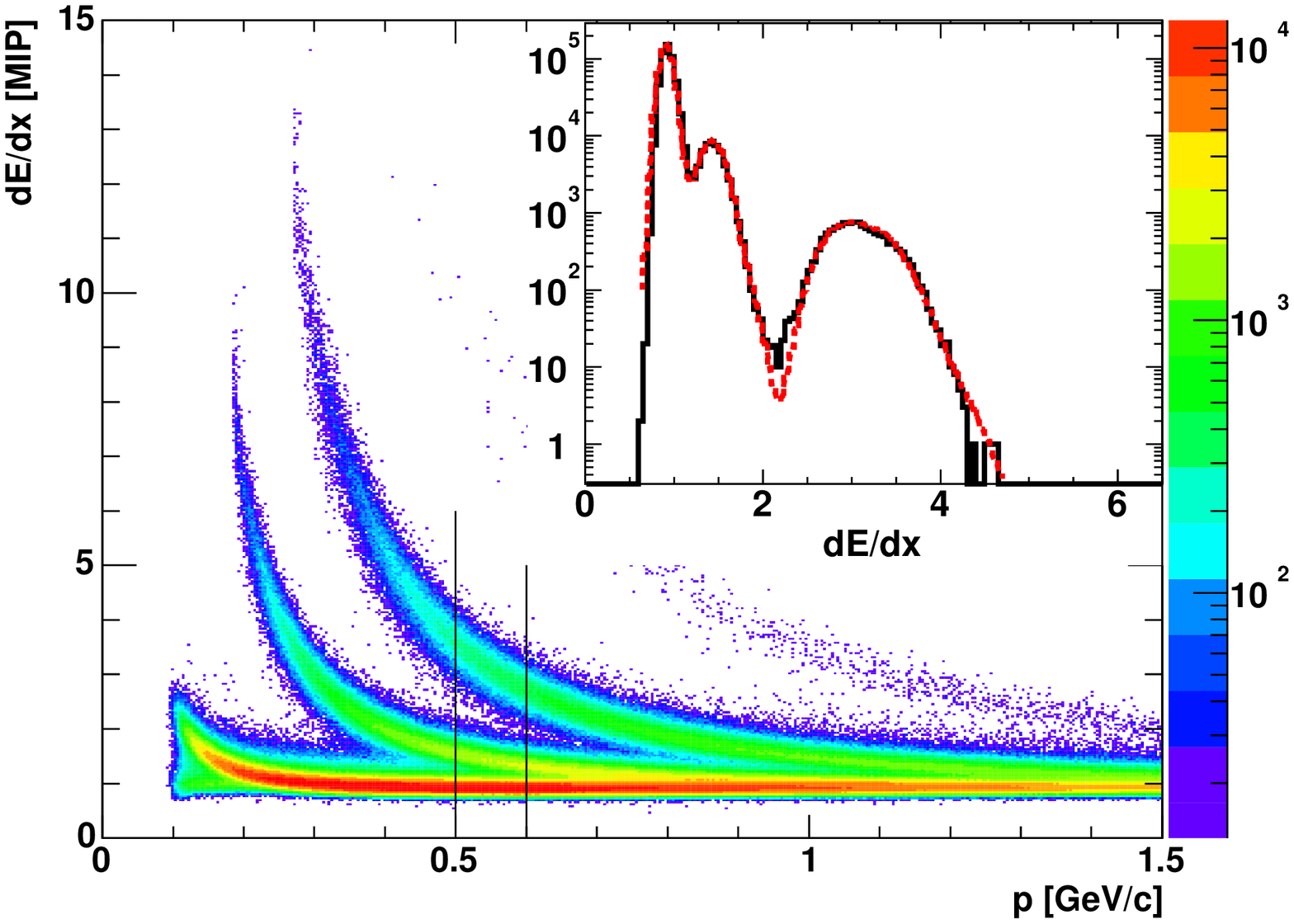}
\includegraphics[width=87mm]{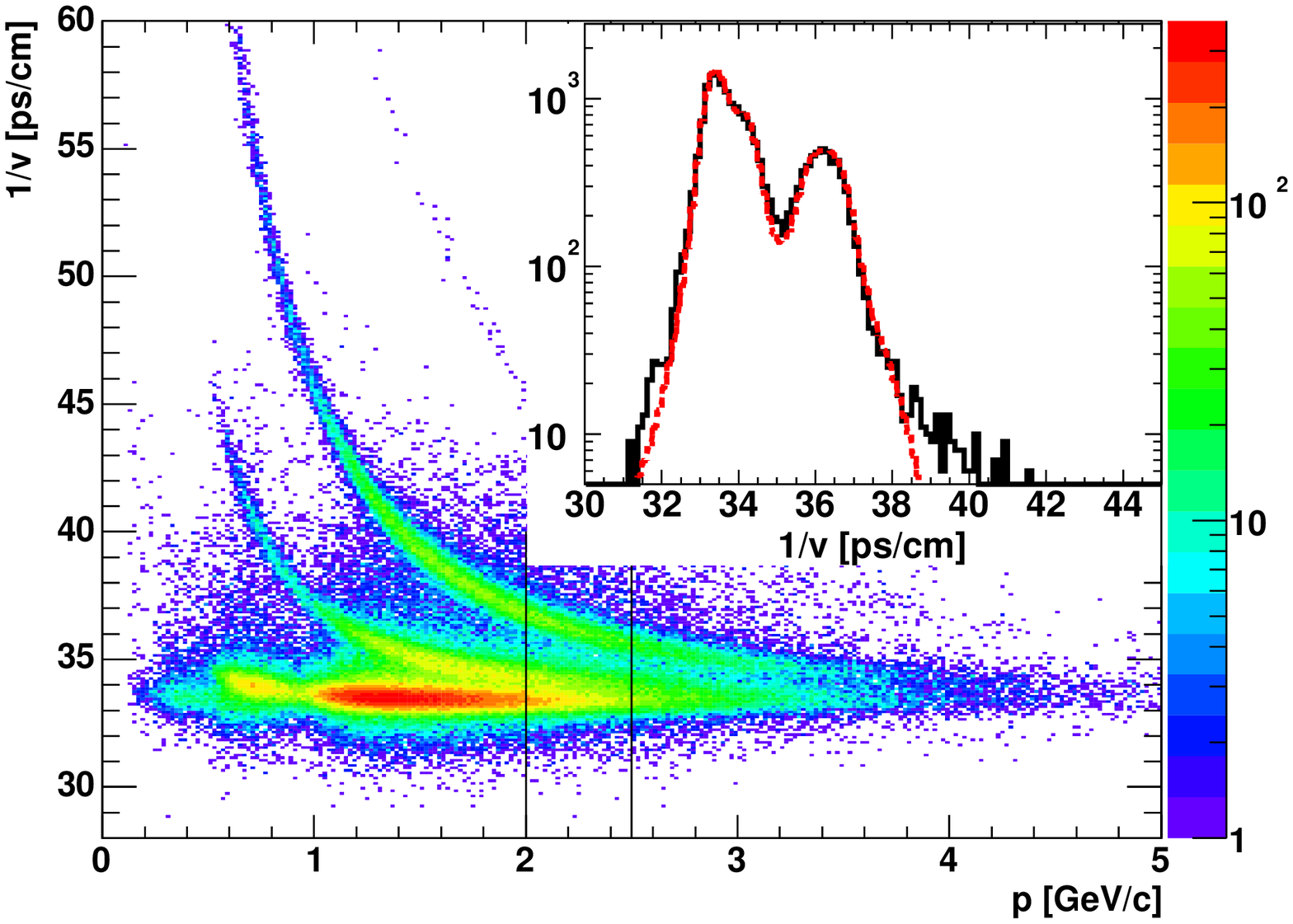}
\caption{\label{pidplot} (Color online) Top panel: $dE/dx$ in the silicon
spectrometer as a function of total momentum. Bottom panel: inverse
velocity of particles in the TOF detector as a function of momentum.
Insets: the spectra are fitted to obtain raw particle yields.
From left to right: pions, kaons and protons are shown (positive charges).
The lines indicate the selected momentum slice to plot the insets.}
\end{figure}


Finally, between 0.5 and 3 GeV/c transverse momentum,
the Time of Flight detector system was used to measure the
velocity. The identification is based on the simultaneous
determination of the velocity and the momentum of these particles.
Details are summarized below.


The magnetic field of the PHOBOS magnet was carefully measured
outside of the dipole gaps (far from the beam line, up to the TOF walls),
as well as modeled with numerical calculations based on the 
electric current and 
detailed coil and magnet yoke geometry and materials. The good agreement 
between measurements and the calculation made it possible to project the 
particle tracks measured by the Spectrometer (where the last Si layer is 
situated at about 1 m distance from the collision point) to the TOF walls 
(at 4-5 m from the collision point) with good precision. 
The hits in the TOF and the charged particle tracks were matched,
and the path lengths of the trajectories were calculated with better than 1 
cm accuracy. The time difference between the TOF and the T0 signal, after 
corrections for cable length differences, drifts caused by temperature
changes, slewing and channel-by-channel delay corrections provided the 
flight time of the particle. The inverse velocity was then calculated for 
each track, and plotted as a function of the particle momentum measured by 
the Spectrometer (shown by the bottom panel of Fig. \ref{pidplot}).

Within a given momentum bin, the trajectories corresponding to different 
species are similar, therefore the path lengths do not vary significantly.
The TOF timing signal, after slewing corrections, is not sensitive to the 
particle mass. The path length error is negligible compared to the timing 
error multiplied by the speed of light.
For these reasons, the inverse velocity ($1/v$) resolution 
is directly proportional to the timing resolution (combined from 
the contributions of the T0 and TOF detectors) and independent of the 
particle mass. 

The $1/v$ distributions in momentum bins were analyzed in a very similar way
to the $dE/dx$ distributions. The line shape for a given species was
a Gaussian complemented with a small tail to account for the non-Gaussian 
nature of the timing error. The mean of the Gaussian was given by 
$1/v=\sqrt{m^2/p^2+1/c^2}$.
In each total momentum bin with finite width, the spread of the velocity 
values according to this formula was taken into account by convoluting
the resolution function with the calculated velocity distribution in the 
momentum bin. The width of the resolution function was constant for all 
species and momenta. The only free fit parameters were the amplitudes 
(particle yields) corresponding to the three hadron species.
The fit functions created this way describe the data well, as shown in the 
inset on the bottom panel of Fig. \ref{pidplot}.

In the high $p_T$ region, where the kaons and pions cannot be 
separated any more, only the meson yield (the sum of kaon and pion yield) 
and the proton yield was measured. For the purpose of constructing the meson line shape, 
the kaon/pion ratio was extrapolated from the lower 
momentum region. For example, the $K^+/\pi^+$ ratio
that was measured to be about 0.65 at $p_T=1.2$ GeV/c was
extrapolated to a range between 0.8 and 2.6 at 3 GeV/c. 
The lack of precise knowledge of that ratio
influences the measured proton yield, causing a small systematic error 
which was estimated to be 2--4\%.

The above fits were performed in total momentum bins for both charge signs
and both magnetic field settings separately.


\section{Corrections}

\begin{figure}[t]
\includegraphics[width=87mm]{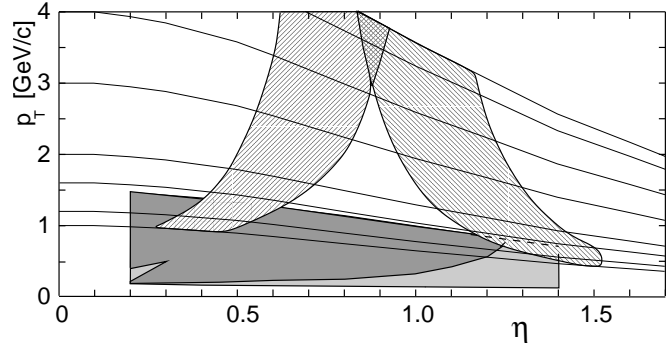}
\caption{\label{acc}
Acceptance of the data used in this analysis in the $(p_T,\eta)$ plane.
The two branches at high $p_T$ represents the TOF, while
the shaded areas at lower $p_T$ represent the Si spectrometer acceptance,
for the two bending directions
(note that the latter for one bending direction {\it contains} the other).
The thin lines are drawn at constant total momentum values.}
\end{figure}

Detector-effects need to be unfolded from the measured distributions in order 
to obtain the true transverse momentum spectra for identified particles.
The various corrections compensate for 
geometrical acceptance and tracking efficiency; occupancy 
in Spectrometer; ghost tracks; momentum resolution; 
feed-down from weak decays of the $\Lambda$ and $\Sigma$ particles;
secondary particles originating from the detector material; and
dead channels in the various detectors.
The details of the above corrections are discussed in Appendix~\ref{appendix_corrections}.

\section{Combining $dE/dx$ and TOF data}


\begin{figure}[t]
\includegraphics[width=82mm]{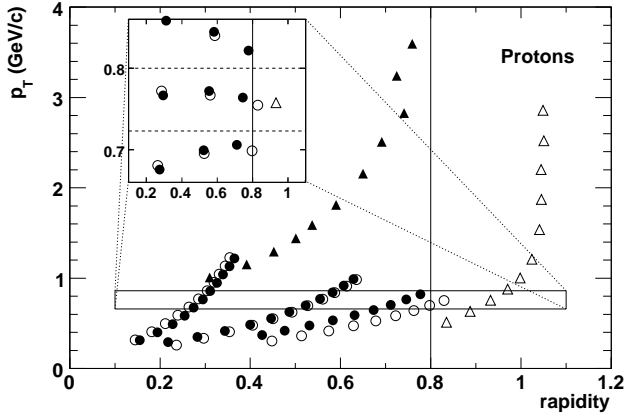}
\caption{Location of PID data-points in $(p_T,y)$-space, for protons from the most central bin. 
Triangles represent TOF, while circles correspond to Spectrometer data that was sliced into three 
$\eta$ bins. Open and closed symbols differentiate between bending directions.
The line at $y=0.8$ indicates the chosen common rapidity value at which all invariant particle yields will 
be evaluated. The inset shows a subset of the same points on a different scale, where the horizontal dashed lines
enclose an example set of data points that will be merged and quoted with a single $p_T$ value.} 
\label{fig:proton_pt_y}
\end{figure}

Because of the dipole magnet configuration and the fixed position of the
TOF wall, the various particle species are detected in slightly different
rapidity regions in the TOF wall, depending on their transverse momentum.
The geometrical acceptance for the identification in the TOF
walls and in the silicon spectrometer for the three species and two magnet
polarities are illustrated in Fig. \ref{acc} in $(p_T,\eta)$-space.
To provide for an easier comparison of our data to theoretical models or other experiments, we synthesize 
the results from these different acceptances to generate the $p_T$ spectrum for each species at a constant 
rapidity.

The Spectrometer data is divided in three bins in pseudorapidity, 
$0.2 \le \eta < 0.6$, $0.6 \le \eta < 1.0$ 
and $1.0 \le \eta \le 1.4$. Particle identification is performed in bins 
of total momentum $p$; the rapidity 
for each species is calculated from the mean $p$ and $p_T$ in each 
(p,$\eta$) bin, and the location of the 
data-points is plotted in $(p_T,y)$-space. An example is shown for 
protons in Figure~\ref{fig:proton_pt_y},
with an inset which details the region around $p_T=0.75$~GeV/c. 
From these plots a common rapidity value is chosen (vertical line at 
$y=0.8$); the invariant yield at this rapidity is evaluated for the whole 
$p_T$ range, using the data measured at different rapidities.

\begin{figure}[t]
\includegraphics[width=82mm]{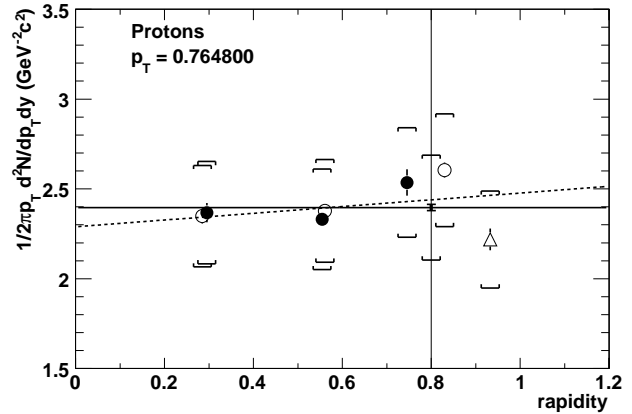}
\caption{Example of the synthesis of TOF and $dE/dx$ data. The invariant
yield  is plotted versus rapidity for protons from central collisions, at
$p_T=0.765$.
The statistical errors on each point are shown by bars, and the brackets
represent the systematic errors. The thick solid and dashed lines show the best constant
and straight-line fits to these points. The vertical line is the common
rapidity value; the brackets on this line represent the total systematic
error on the synthesized invariant yield.}
\label{fig:synthesis_1}
\end{figure}

The procedure to synthesize all of the data for a particular species into
a $p_T$ spectrum at $y=0.8$ begins by choosing data-points which are close in $p_T$. 
An example of such a small $p_T$ region is shown in the inset of 
Figure~\ref{fig:proton_pt_y} by the two horizontal dashed lines. The seven data 
points falling between the lines will be combined in the following way.
The invariant yield, 
$Y=\frac{1}{2\pi p_T}\frac{d^2N}{dp_T dy}$, 
of all points in the selected region is plotted versus rapidity
(see Fig.~\ref{fig:synthesis_1}).
The simplest assumption is that the invariant yield is constant over the 
measured narrow rapidity interval near mid-rapidity.
We therefore take the best constant fit to these points as the value of the invariant yield at this $p_T$. 
For comparison, we also fit a straight line to the points; the difference between the constant fit and 
this straight line evaluated at the common rapidity point is taken as a measure of the systematic error 
introduced by this assumption.
This process is illustrated for a single $p_T$ `slice' in Figure~\ref{fig:synthesis_1}.
The statistical error on the synthesized invariant yield is the propagated error from the 
individual points (shown by small horizontal bars).

An important consideration in finding the yields which are fit as shown 
in Figure~\ref{fig:synthesis_1} is
the fact that, although the data-points have similar $p_T$ values, they are not identical. Therefore, 
it is necessary to account for the strong $p_T$ dependence of the particle yields. 
This is done using a 
Taylor expansion:
\begin{equation}
 f(p_T + \Delta p_T) \approx f(p_T) + \Delta p_T \frac{df(p_T)}{dp_T}
\end{equation}
where $f(p_T)$ is the yield of an individual data point and $\Delta p_T$ is the difference between the $p_T$ 
of that point and the average $p_T$ of all points being combined. 
The slope of the $p_T$-dependence, $\frac{df(p_T)}{dp_T}$, is found by doing a fit
to all of the raw data points for a particular bending direction (i.e. either the open or closed 
symbols in Figure~\ref{fig:proton_pt_y}), ignoring for the moment that each of these points is
at a different rapidity.
The yield of each individual data point is `perturbed' in this way to a common $p_T$ value before 
they are merged to find the yield at $y=0.8$.
In principle, this adjustment of the data points could be iterated using the spectrum found 
after projecting to $y=0.8$ in order to obtain a more precise value of the slope, $\frac{df(p_T)}{dp_T}$.  
In practice, however,  
the applied shifts were so small that such a refinement was unnecessary.

\section{Integrating the ${\rm p_T}$ Spectra}

Proton and antiproton $d^2N/dp_Tdy$ transverse momentum distributions were integrated to obtain the total 
particle yield $dN/dy$.
We integrate over the measured data points and extrapolate over the unmeasured low-$p_T$ region. Because the 
$p_T$ spectrum falls so sharply, the high-$p_T$ region beyond the measured points makes a negligible 
contribution to the total yield and is not included. The low-$p_T$ extrapolation uses a simple straight line 
from zero to the first data point; this result is compared to that obtained using a variety of 
physically-motivated fit functions.

Statistical errors on the integral turn out to be negligible.
Systematic errors on the total yield come from propagation of the errors 
on the individual data-points, plus additional uncertainty which arises 
as a result of the extrapolation over the unmeasured low-$p_T$ region.



\section{Results}

\begin{figure}[t] 
\includegraphics[width=87mm]{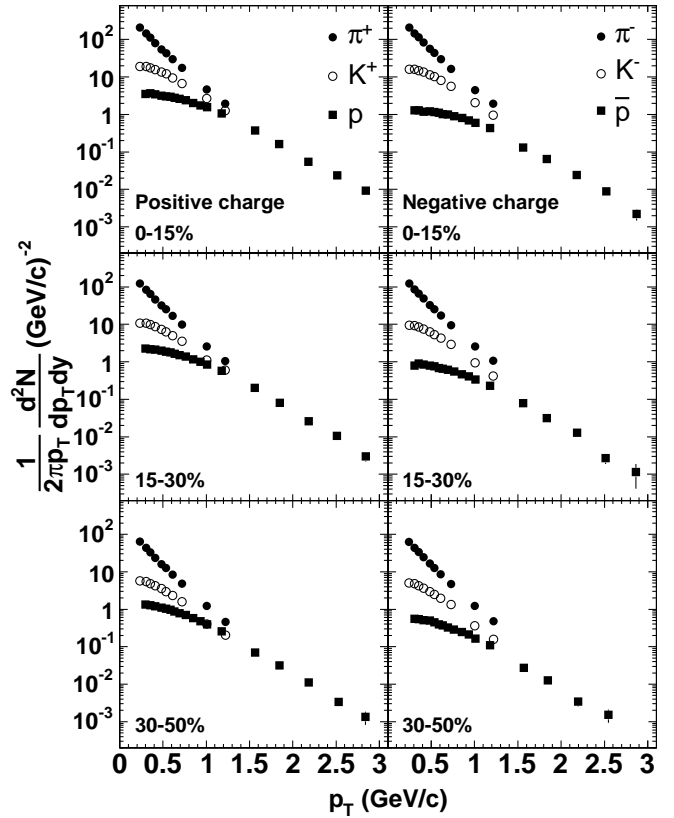}
\caption{\label{spectra}
Transverse momentum spectra of $\pi^+$, $\pi^-$, $K^+$, $K^-$,
$p$ and $\overline{p}$ particles for the three centrality classes in Au+Au collisions at
$\sqrt{s_{_{NN}}}$=62.4 GeV. Left panels: positively-charged particles, right panels: negatively-charged particles.}
\end{figure}

After all corrections are applied, the 
invariant yields ($d^2N/2\pi p_T dp_T dy$) of each species in the three centrality classes are 
plotted as a function of transverse momentum in Fig.~\ref{spectra}, 
for Au+Au collisions at $\sqrt{s_{_{\rm NN}}}=62.4$ GeV. 
As described in the previous section, the final data is extracted at $y=0.8$ 
for all species, by combining the actual measurements 
from the $dE/dx$ and TOF identification methods,
which cover the geometrical acceptance shown in Fig.~\ref{acc}.

Above a transverse momentum of 1.4~GeV/c, kaon yields cannot be 
distinguished from pions.
The proton and antiproton yield can be measured up to $p_T=3$~GeV/c,
where the limit is governed by the statistics of the data sample.

Systematic uncertainties on the invariant yields arise from various sources.
The efficiency of track reconstruction and the acceptance of the 
Spectrometer carries a systematic error between 5 and 9\%, decreasing 
with $p_T$. 
The feed-down correction for protons is estimated with a 4-8\% 
precision, and this uncertainty also decreases at larger transverse momenta.
There are several smaller error sources such as the feed-down to pions (1\%),
secondaries contributing to the proton yield (1\%), dead channels in the 
Spectrometer (5\%) and the TOF (2\%), reconstructed ghosts/fakes (2\%),
the occupancy correction in the Spectrometer (3\%) and the systematic 
error from the procedure to combine the data sets from different detectors 
(4-5\%). Overall, the systematic uncertainty on the proton and antiproton yield 
is 14-16\%, while for pions it is 13-15\%, and 11-14\% for kaons.
For all species, the uncertainties decrease slightly with increasing 
transverse momentum out to $p_T\approx 2$~GeV/c and then rise slowly.

Only a mild centrality dependence can be observed in the data, while the
difference between the shapes of the $p_T$ spectra of various species 
is significant. At high transverse momenta, the proton and 
antiproton spectra become closely exponential in the measured $p_T$
range, while they flatten out at low $p_T$. 

\begin{figure*}[ht]
\includegraphics[height=53.9mm]{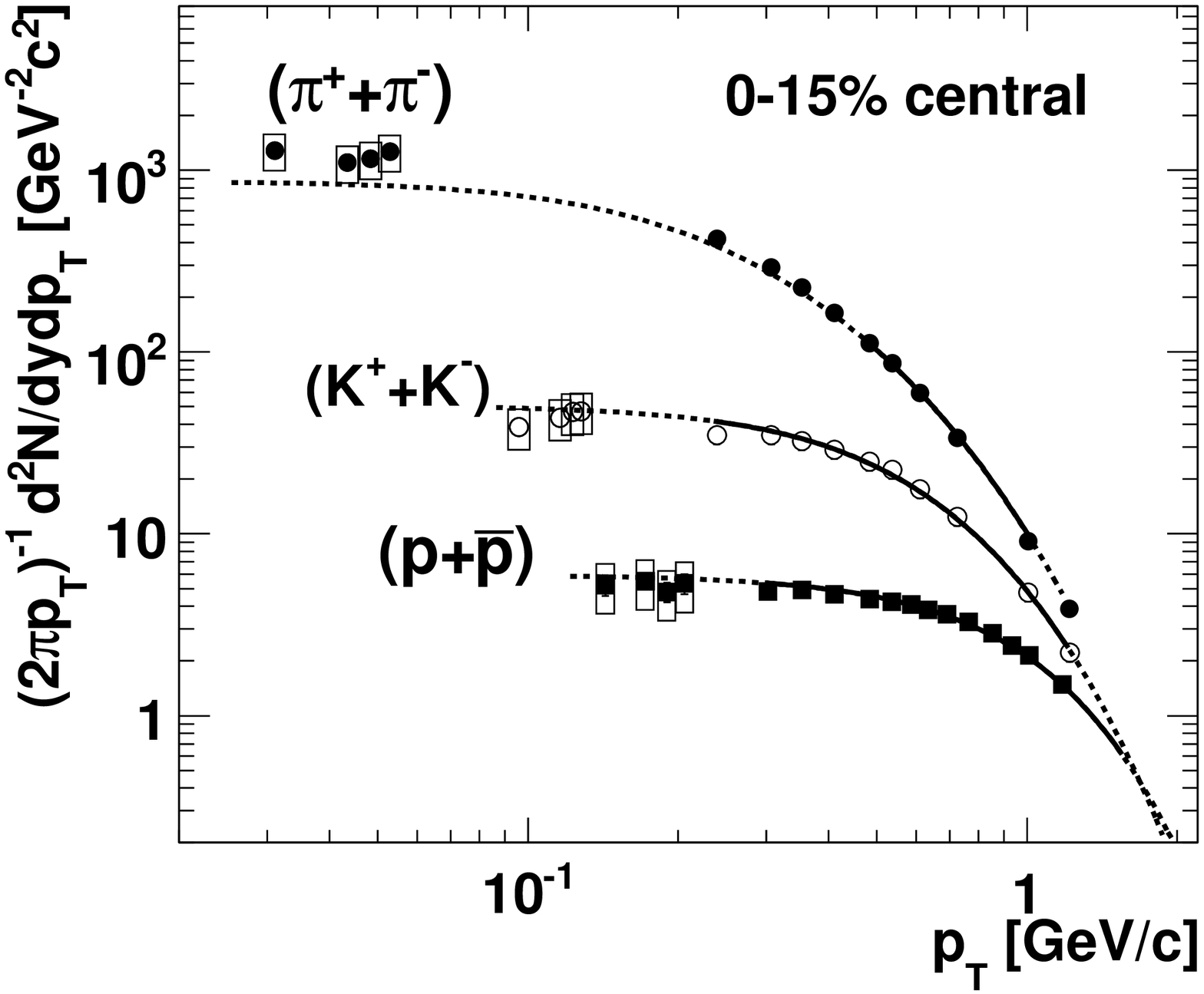}
\includegraphics[height=53.9mm]{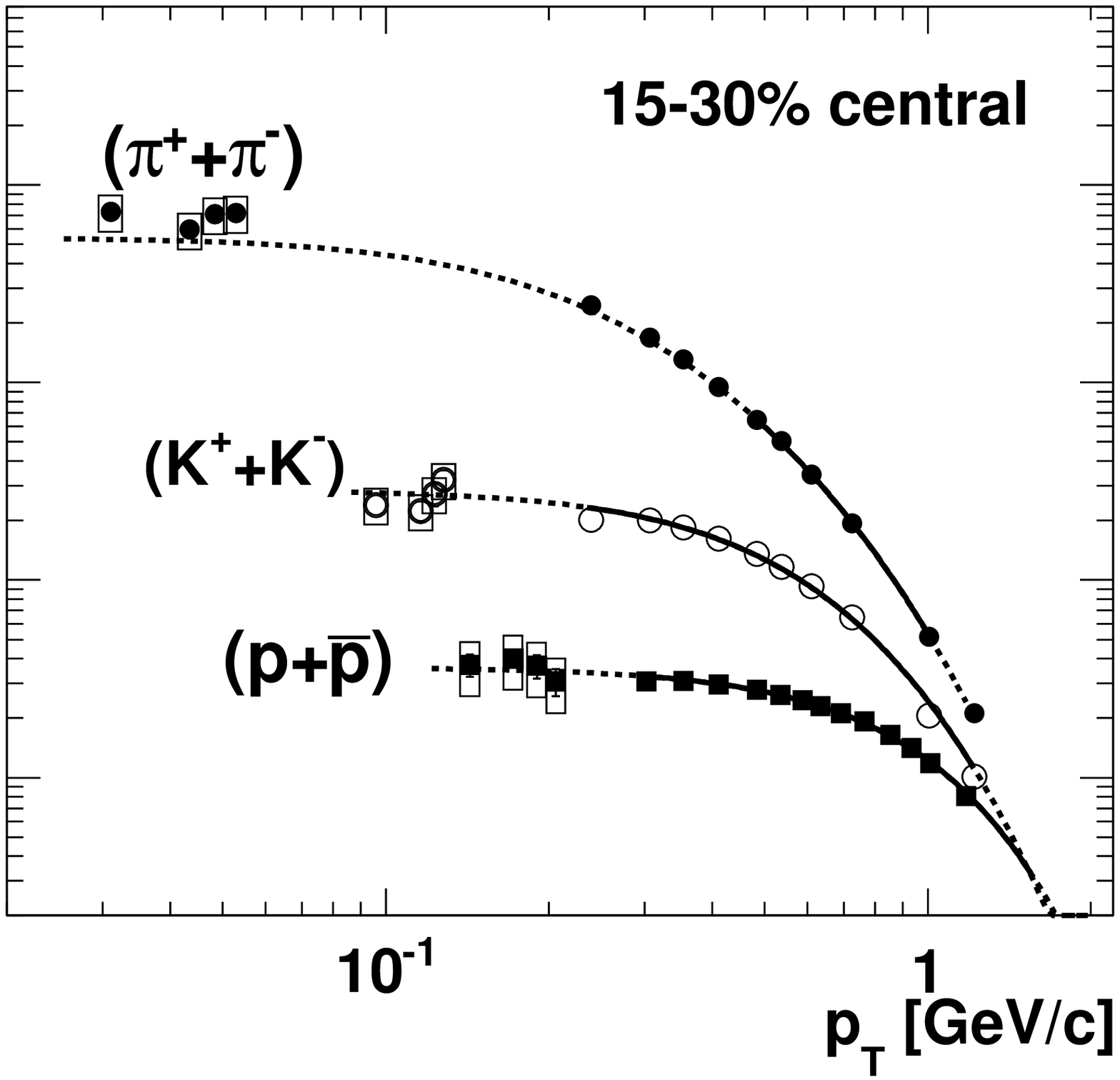}
\includegraphics[height=53.9mm]{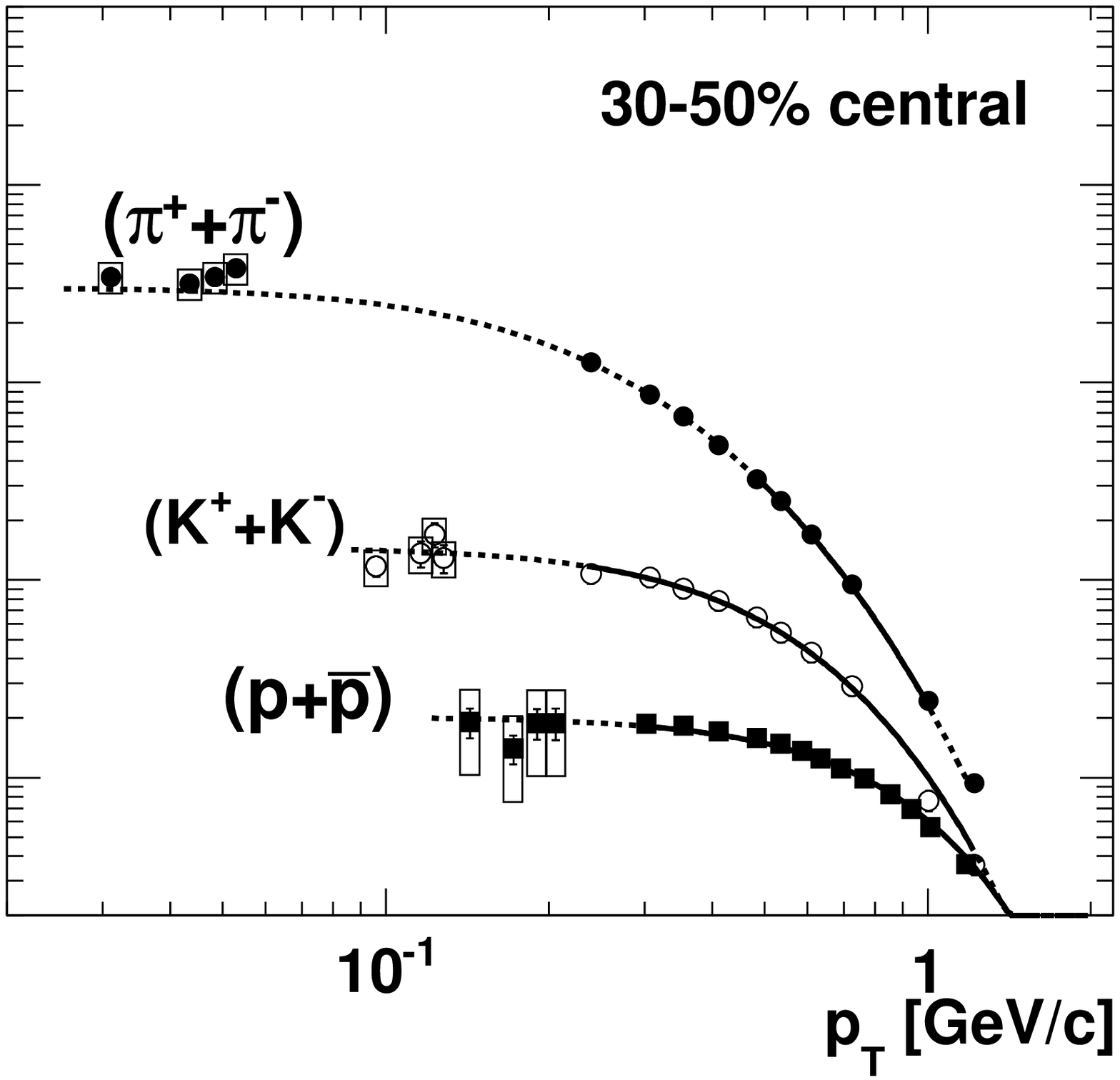}
\caption{\label{bwfit}
Blast-wave fits to identified 
particle spectra in Au+Au collisions at $\sqrt{s_{_{NN}}}=62.4$ GeV.
Error bars represent statistical errors. Systematic errors are shown by 
the boxes for the data from the very low-$p_T$ analysis.}
\end{figure*}

Figure~\ref{bwfit} shows the comparison of these intermediate $p_T$ identified 
particle yields with the data at very low $p_T$, obtained from the
analysis of particles stopping in the 
5th active Si layer of the Spectrometer.
The yields corresponding to the two charge signs for a particle were added here,
since in the very low $p_T$ analysis it is not possible to determine the charge of the particle.
A fit to the data was performed using a
blast-wave parametrization \cite{Schnedermann:1993ws}.
Each of $\pi^+$, $\pi^-$, $K^+$, $K^-$, $p$ and $\overline{p}$ were fit separately, and then the fit functions were summed over charge sign.
Pions below 0.45~GeV/c were not used in the fits to exclude
products of strongly decaying resonances. The data points from the
very low-$p_T$ analysis were also excluded from the fit; the fit obtained at intermediate $p_T$ is then extrapolated down and compared to these data.
Based on the high quality of the fit, one can conclude that a radial 
expansion, characterized by a radial flow velocity of about $\beta=0.75$,
describes the data well, over the full transverse momentum range 
studied here. 
No anomalous enhancement of invariant pion yield
at very low $p_T$ is observed, when compared to a simple expectation
involving radial expansion, similar to what was seen for the very low 
$p_T$ results obtained for Au+Au collisions at $\sqrt{s_{_{NN}}} =$ 
200~GeV in \cite{phobos:stopping}.
This statement is also valid for the other particle species.
The slight excess we observe in pion yields compared to the blast-wave parametrization is 
explained by the fact that the fit does not incorporate products of strong resonance decays, and those decays contribute significantly 
 to the pion yield at low $p_T$.

The blast-wave parameters (the velocity parameter $\beta$ and the
temperature parameter $T$) appear to be very similar for different centrality 
bins: $T$=103, 102, 101~MeV and $\beta$= 0.78, 0.76, 0.72 for the 
central (0-15\%), mid-peripheral (15-30\%) and most peripheral (30-50\%) 
data sample, respectively. By including the very low-$p_T$ proton and kaon 
data points, these parameters change by less than 6 MeV and 0.02, 
respectively.

\begin{figure}[t]
\includegraphics[width=77mm]{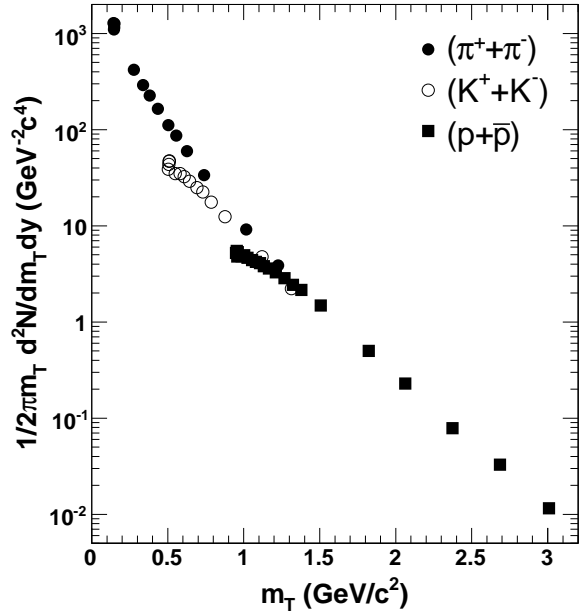}
\caption{\label{mtspectra}  
Transverse mass spectra of pions, kaons and protons in Au+Au collisions at
$\sqrt{s_{_{NN}}}$=62.4 GeV energy. Particle and antiparticle yields
are added to accommodate all three identification methods.
Statistical errors are small compared to the
symbol size.}
\end{figure}

It was shown earlier \cite{Veres:2004kc}
that the transverse mass ($m_T=\sqrt{m^2+p_T^2}$) spectra of
the various hadron species in d+Au collisions at $\sqrt{s_{_{NN}}}$=200 
GeV energy satisfy a scaling 
law within the experimental errors: they differ only by an 
$m_T$-independent scale factor. 
However, the scaling was shown to be violated in Au+Au collisions at 
the same collision energy \cite{phobos:stopping,Adler:2003cb}. 
Figure \ref{mtspectra} shows that the scaling violation is similar
at $\sqrt{s_{_{NN}}}$=62.4 GeV in Au+Au collisions.
Another important observation is that in Au+Au 
collisions the invariant yields for the three species seem to 
approximately converge for the $m_T$ region from 1 to 1.5 GeV/$c^2$,
while in d+Au collisions at 200 GeV the 
kaon yield is suppressed with respect to the pion and proton yield
by about a factor of two in the same $m_T$ range \cite{Veres:2004kc}.

Since the pion spectrum falls faster with transverse momentum than the 
proton spectrum (see Fig.~\ref{spectra}), 
proton yields dominate over mesons
at high $p_T$. This phenomenon may carry an important physics   
message about the relevant degrees of freedom in the radially expanding   
medium created by heavy-ion collisions. A similar observation was made previously at 200 GeV
collision energy \cite{Adler:2003cb}, where antiprotons were also found to dominate over negatively-charged mesons.

\begin{figure}[t]
\includegraphics[width=87mm]{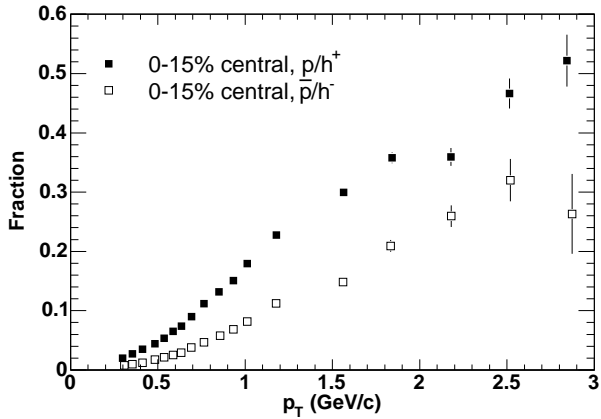}
\caption{\label{protonfractions}
Fraction of protons among all positive hadrons, and antiprotons among
all negative hadrons in Au+Au collisions at
$\sqrt{s_{_{NN}}}$=62.4 GeV.}
\end{figure} 

In order to study baryon dominance over mesons at high $p_T$, the fraction 
of protons among all positive hadrons ($p/h^+$) and the 
fraction of antiprotons among all negative hadrons ($\overline{p}/h^-$) is plotted in Fig.~\ref{protonfractions}, for 
central Au+Au collisions at 62.4 GeV. 
This particular ratio was chosen because it can be measured up to 3~GeV/c
transverse momentum 
(by taking the $h^+$ yield as the sum of $p$, $\pi^+$ and $K^+$ yields, 
etc.), while the measurable $p_T$ range of the $p/\pi^+$ and 
$\overline{p}/\pi^-$ ratios extends only up to 1.4 GeV/c.

The $p/h^+$ ratio reaches 0.5 above $p_T$=2.5~GeV/c, which means that the 
proton yield becomes comparable to the sum of pion and kaon yields: 
$Y_p\approx Y_{\pi^+}+Y_{K^+}$. The $\overline{p}/h^-$ ratio reaches 1/3 
at around the same $p_T$ value, which means that either 
$Y_{\overline{p}}\geq Y_{\pi^-}$ or $Y_{\overline{p}}\geq Y_{K^-}$ becomes 
true, depending on whether the $K^-/\pi^-$ ratio is above or below unity
(we were not able to measure this above  $p_T$=1.4 GeV/c).
One can conclude that baryons become the dominant particle species at 
about 2.5$-$3.0 GeV/c transverse momentum in central Au+Au collisions at 
62.4 GeV energy.

\begin{figure}[t]
\includegraphics[width=85mm]{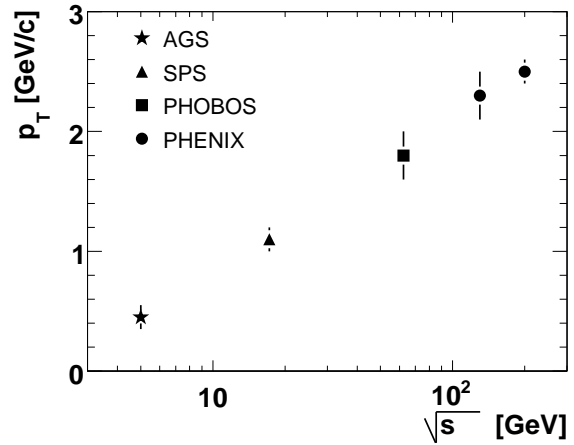}
\caption{\label{crossing}
The $p_T$ value where the $\pi^+$ and proton invariant yields
become equal in central Au+Au (Pb+Pb) collisions, as a function of
$\sqrt{s_{_{NN}}}$. All data have been corrected for feed-down from weak 
decays. See text for references.}
\end{figure}

Fig.~\ref{crossing} illustrates the evolution of this baryon dominance with collision energy.
For central heavy-ion (Au+Au or Pb+Pb) collisions at AGS, SPS and RHIC energies, we study the $p_T$ value at which the invariant yields of protons and 
positive pions at mid-rapidity become equal.
Data (fraction of most central events) are taken from the experiments:
E802~\cite{Ahle:1998jc}~(4\%),
NA44~\cite{Bearden:2002ib}~(3.7\%),
NA49~\cite{Anticic:2004yj}~(5\%),
PHENIX~\cite{Adler:2003cb,Adcox:2003nr}~(5\%)
and PHOBOS~(15\%).
Although the pion spectrum from the present analysis
does not reach above 1.4 GeV/c transverse momentum,
the meson yield ($K+\pi$) can be measured up to 3 GeV/c.
By allowing the extrapolation of the $K^+/\pi^+$ ratio to 2 GeV 
to change within reasonable limits, the location of the `crossing point' can
still be estimated with a meaningful systematic error.
A remarkably smooth collision energy dependence of the `crossing' $p_T$ 
value is observed. At low energies, the abundance of produced pions is 
naturally low compared to high collision energies, while baryon number 
conservation ensures that a significant fraction of the large number of 
initial state protons are found in the final state, thus the invariant 
yields of protons and positive pions become comparable already at low 
$p_T$. With increasing energy, this $p_T$ value grows, mainly due to the  
approximately logarithmically increasing number of produced pions.

The dominance of the baryon yield is perhaps more interesting for 
antiparticles:
antiprotons and negative pions, since all antiprotons are newly produced 
in the collision. Because of the strongly energy dependent production 
cross section of antiprotons, for the antiproton yield and the $\pi^-$ 
yield to be comparable, the highest RHIC collision energies are 
needed. This was observed for the first time in Au+Au collisions at 
$\sqrt{s_{_{NN}}}=130-200$~GeV~\cite{Adcox:2001mf}. 

It is well known that the antiproton to proton 
ratio at midrapidity in Au+Au (Pb+Pb) collisions increases 
from a very small value to almost unity within the collision energy 
range plotted in Fig.~\ref{crossing}.
If one makes the simple assumption that part of the 
protons are transported from beam rapidities to midrapidity while 
other protons are pair produced (in approximately the same amount as the 
antiprotons), this ratio is the fraction of newly produced protons 
among all protons. At low energies, almost all midrapidity protons are 
transported from beam rapidities, while at the highest RHIC energy almost 
all protons are pair produced according to the above simple picture.
Therefore, the fact that the proton and $\pi^+$ yields become comparable 
to each other at high $p_T$ also has implications on the 
baryon/meson differences of particle production at midrapidity.
In jet fragmentation, which should be an important mechanism at high 
$p_T$, the expectation is to have a small baryon/meson ratio, as 
observed in elementary particle collisions \cite{Abreu:2000nw}. 
However, if a quark recombination process is dominant in the creation of 
baryons and mesons at midrapidity, the observed high baryon/meson ratio at 
$p_T$ of a few GeV/c is expected \cite{Hwa:2002tu}.

\begin{figure}[t]
\includegraphics[width=87mm]{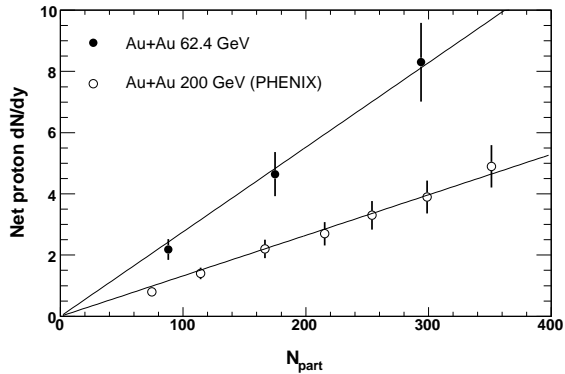}
\caption{\label{netprotons}
Net proton ($p-\overline{p}$) yield at midrapidity, as a function of the
number of participant nucleons in the collision. Filled symbols are Au+Au
collisions
at 62.4 GeV energy from the present analysis, open symbols are Au+Au
collisions at 200 GeV from PHENIX \cite{Adler:2003cb}.
In the latter case, the experimental errors on the proton and antiproton
spectra were assumed to be completely correlated.
The proportionality lines are only to guide the eye.
}
\end{figure}

It is worthwhile to discuss the baryon production and baryon transport 
features in connection with the above observations. The proton and 
antiproton invariant yields were integrated over $p_T$ at midrapidity, and 
subtracted to evaluate the net proton (defined as $p-\overline{p}$) 
rapidity density ($dN/dy$). 
The proton, antiproton and net proton integrated yields for all centrality 
bins are given in Table~\ref{table:proton_dndy}; the net proton yields 
are plotted in Fig.~\ref{netprotons} as a function of the number of 
participant nucleons in the collision. For comparison, data 
measured by the PHENIX experiment at 200 GeV~\cite{Adler:2003cb} are also 
shown, where the experimental errors on the proton and antiproton 
spectra are assumed to be completely correlated. Thus, the 
errors assigned to the net proton yields are lower limits.
At both energies, the net proton density at midrapidity 
appears to be closely proportional to the number of participant nucleons.
That proportionality means that the number of protons transported to 
midrapidity per initial state participant does not depend on the 
average number of collisions a participant suffers in the collision (which 
is strongly centrality dependent, changing between 2.91 and 4.65 for the 
62.4 GeV data, and between 3.23 and 6.07 for the 200 GeV data plotted in 
Fig.~\ref{netprotons}). 

The $\overline{p}/p$ ratios in heavy ion collisions measured extensively 
by other experiments
(e.g. \cite{Adler:2003cb} and \cite{Bearden:2003fw}
at 200 GeV and \cite{Adler:2001bp} at 130 GeV) 
are approximately centrality-independent, and also 
approximately $p_T$-independent below 4-5 GeV/c transverse momentum.
This means that the midrapidity proton and antiproton yields (and not only 
their difference) are approximately proportional to the number of 
participants. 

\section{Summary}

The first identified hadron transverse momentum spectra
from the Au+Au run at 62.4 GeV collision energy at RHIC were presented.
This is also the first identified transverse momentum spectra 
analysis using the PHOBOS silicon Spectrometer and the first analysis 
using the Time of Flight detector. The very low $p_T$ data points are 
unique at RHIC. 
The identified particle spectra measured at 62.4 GeV also bridge a 
significant gap in collision energy between heavy ion collision data 
taken at the highest SPS energy and the high RHIC energies.

\begin{table}[t]
\begin{ruledtabular}
\begin{tabular}{cccc}
Centrality & $\frac{dN}{dy}$($p$) & $\frac{dN}{dy}$($\bar{p}$) & $\frac{dN}{dy}$($p-\bar{p}$) \\
\hline
0--15 \% & $13.4 \pm 1.9$ & $5.1 \pm 0.7$ & $8.3 \pm 1.3$ \\
15--35 \% & $7.7 \pm 1.1$ & $3.0 \pm 0.4$ & $4.6 \pm 0.7$ \\
35--50 \% & $3.8 \pm 0.5$ & $1.6 \pm 0.2$ & $2.2 \pm 0.3$ \\
\end{tabular}
\end{ruledtabular}
\caption{Proton, antiproton and net proton ($p-\bar{p}$) integrated yields.
The errors quoted are systematic, statistical errors are negligible.
Correlations in the systematic errors for protons and antiprotons are
included in the systematic error on the net proton yield.}
\label{table:proton_dndy}
\end{table}

Invariant cross sections of protons, antiprotons, pions and 
kaons in three centrality bins were presented. No significant enhancement
of pions was found compared to a simple blast-wave parametrization
at very low $p_T$, and the transverse momentum spectra are consistent with 
radial expansion down to very low $p_T$. 

Remarkably, at about 2.5$-$3 GeV/c transverse momentum, the proton and 
antiproton yields exceed the yield of at least one of the meson species.
The energy dependence of this baryon dominance was studied in detail.
It was shown that the transverse momentum value where the $p/\pi^+$
ratio exceeds unity in central Au+Au (Pb+Pb) collisions is a very smooth 
function of the collision energy. The net baryon yield at midrapidity was, 
surprisingly, found to be closely proportional to the number of 
participant nucleons in the collision. 

\begin{acknowledgments}
%
%
%
%
This work was partially supported by U.S. DOE grants 
DE-AC02-98CH10886,
DE-FG02-93ER40802, 
DE-FC02-94ER40818,  
DE-FG02-94ER40865, 
DE-FG02-99ER41099, and
W-31-109-ENG-38, by U.S. 
NSF grants 9603486, 
0072204,            
and 0245011,        
by Polish KBN grant 1-P03B-062-27(2004-2007),
by NSC of Taiwan Contract NSC 89-2112-M-008-024, and
by Hungarian OTKA grant (F 049823).
\end{acknowledgments}
\appendix

\section{Event Selection and Centrality}
\label{appendix_evsel}

Determining the centrality of heavy ion collisions is important for 
event characterization. The centers of the ions travel along parallel 
lines before collision, displaced by $b$, called the impact parameter.
The number of nucleons `participating' (colliding at least with one other 
nucleon) in the collision, $\rm N_{\rm part}$, and the number of binary 
collisions between nucleon pairs, $\rm N_{\rm coll}$, can be connected to
the impact parameter, using a simple geometrical model 
calculation.

None of these three quantities can be measured directly,
but $\rm N_{\rm part}$ and $\rm N_{\rm coll}$ can be estimated
from measured quantities. In PHOBOS, charged particle multiplicities in 
various pseudorapidity ($\eta$) regions are used to estimate 
centrality \cite{phobos:white}.
These $\rm N_{\rm part}$ and $\rm N_{\rm coll}$ centrality measures 
make it possible to study scaling of the hadron yields in peripheral and 
central heavy ion collisions, compared to elementary (p+p) collisions.
The expectation is, that the
number of `hard' parton scatterings with large momentum transfer and small 
cross section should be proportional to $\rm N_{\rm coll}$.  However, it was 
observed that the total number of charged particles as well as 
the hadron yields at low and even high $p_T$ is 
(approximately) proportional to $\rm N_{\rm part}$. It is important to 
study whether these observations hold for various hadron species 
separately.

The initial event selection used the time difference of the Paddle and 
Zero Degree Calorimeter signals, which were required to be less than 4 ns, 
to exclude beam-gas interactions. For very 
central events with few spectator neutrons but 
high Paddle signals, a hit in the ZDCs was not required. 
Separate logic filtered out events happening shortly after or before 
another collision, to avoid pile-up.

Since the geometrical acceptance of the Spectrometer strongly depends on 
the vertex position in the beam direction, an optimal vertex range was selected using the 
difference of the time signals from the two T0 detectors online,
and a narrower, 20 cm wide vertex region was selected later offline, 
based on the precise vertex reconstruction using the silicon detectors.

The efficiency of the above trigger and selection was measured by 
comparison to a minimum bias type trigger, which required only a single hit 
in both Paddle detectors. Since the minimum bias trigger was 
only $91\pm2\%$ efficient, a full Monte Carlo simulation accounted for the 
loss of the most peripheral events. After the efficiency determination, the 
data with the restrictive trigger could be sorted into bins according to 
percentage fractions of the total inelastic cross section, where the 
experimental variable measuring centrality was chosen to be the 
total energy deposited in the Paddles.

In the present analysis, the most central 50\% of the total cross section 
was used, where the triggering and vertex finding was still fully 
efficient, and divided into the three centrality classes: the 15\% most 
central events, and two more bins between 15--30\% and 30--50\%.  

Comprehensive Monte Carlo simulations of the Paddle signals, 
including Glauber model calculations of the collision
geometry, allowed the estimation of the number of participating nucleons,
$\langle {\rm N_{part}} \rangle$, for each total cross section bin.
The systematic uncertainty of ${\rm N_{part}}$ was estimated with MC 
simulations taking into account possible errors in the overall trigger 
efficiency and by using different types of event generators.

The average number of binary nucleon-nucleon collisions, 
$\langle \rm{N_{coll}} \rangle$, was found from a parametrization of the 
results of a Glauber calculation \cite{phobos:white}: 
$\langle \rm{N_{coll}} \rangle$=
0.296$\langle \rm{N_{part}} \rangle^{1.376}$.
In this way, the $\rm{N_{part}}$ determination, which is relatively 
insensitive to the parameters of the MC event generators, and the 
relation between $\rm{N_{coll}}$ and $\rm{N_{part}}$, which depends 
strongly on cross sections used, are conveniently decoupled.



\section{Track Reconstruction}
\label{appendix_tracking}

Signals in individual Spectrometer pixels are first clustered to obtain `hits'.
The thresholds are determined in terms of the energy deposited by a Minimum 
Ionising Particle (MIP), which is found to be 80~keV for the
300~$\mu$m-thick PHOBOS silicon sensors.
Any pixel with a signal above 0.15~MIP is considered a candidate for merging.
Adjacent pads in the horizontal direction can be merged if they are above 
this threshold, up to a limit of 8 pads per hit candidate.
A set of merged pads must sum to over 0.5~MIP before being declared a `hit'.
There is no vertical merging of pixels.

 Straight-line tracks are reconstructed in the low-field region of the 
 Spectrometer (in the layers closest to the beam pipe) using a 
 road-following algorithm. Track candidates are 
 required to have hits from at least five of the six inner layers and to 
 point back to the independently-determined event vertex. 

 Combined with the event vertex location, a pair of hits from consecutive layers
 in the high field region  can be mapped to the total momentum $p$ and 
polar angle $\theta$ of the track which would have produced those hits.
 Track reconstruction therefore looks for clusters of hit pairs in 
 $(p,\theta)$ space that correspond to a trajectory which traverses all 
 layers.

 Curved and straight track pieces are joined by requiring consistency in 
 their $\theta$ and $\phi$ angles and in the average energy deposited per 
 hit. 

 The track trajectory is found by propagating the momentum vector through 
 the  PHOBOS magnetic field using a Runge-Kutta algorithm and detailed 
 field map. At this stage, no effects such as multiple scattering or 
 energy loss are included. 
 This determines the trajectory relative to which hit residuals are 
 calculated.

 The track momentum is determined by a $\chi^2$ fit to the reconstructed 
 hit pattern. The error for each hit position is influenced by deflection  
 of the charged particle due to multiple scattering and pixelisation of 
 the silicon sensors.
 Multiple scattering also introduces a correlation between the errors on 
 different hits, because a deflection in one layer tends to produce 
 systematic offsets in the hit-residuals for subsequent layers.

 The PHOBOS tracking package uses a complete covariance matrix to account 
 for these correlations. Covariance matrices are pre-generated using Monte 
 Carlo simulation of pions passing through the Spectrometer, including the 
 effects of multiple scattering and energy loss due to interaction with 
 the detector material.

 The actual minimisation procedure to find the best trajectory uses 
 a standard `downhill simplex' technique \cite{numrec:num_recipes}. 
 This  multi-dimensional minimisation method does not require knowledge of the 
 derivatives of the function being minimised. That is important
 because the non-uniform magnetic field means that there is no 
 analytic form for the particle trajectory and, therefore, accurate derivatives cannot 
 easily be computed.
 The procedure not only finds the best-fit trajectory but also assigns a 
 goodness-of-fit value. This variable is found to be a powerful tool 
 for rejecting tracks resulting from incorrect hit associations. 
 Distinct tracks are not allowed to share more than two hits. If a pair 
 of candidates share more than two hits, the one with the lower 
 fit probability is discarded.

 Since the track reconstruction procedure assumes pions, heavier particles 
 like kaons and protons will deposit more energy than the template track 
 and hence be assigned a reconstructed momentum that is systematically too 
 small.
 Monte Carlo simulations are used to obtain momentum correction factors 
 for tracks which are identified to be kaons or protons.

 The track reconstruction efficiency is about 80\%.
 The momentum resolution achieved  by this tracking package  is about 1\% 
 for total momentum $p \approx $1~GeV/c and rises linearly with $p$, but is 
 still less than 5\% for $p$=8~GeV/c.


For particle identification by Time-of-Flight, the reconstructed Spectrometer
tracks are extrapolated towards the TOF detector. This extrapolation uses the
same Runge-Kutta algorithm as the track reconstruction, in combination with
a detailed map of the (small but non-zero) magnetic field in the region between
the Spectrometer and the TOF. 
It is found that a better match is obtained when the track extrapolation uses
the momentum value obtained by fitting \emph{without} covariance matrices -- 
this technique gives a better fit to the track momentum as it exits the
Spectrometer.
However, the original momentum value is retained and is still the value used 
for physics.

Signals in the TOF sensors are checked for good timing characteristics and
sufficient deposited energy to indicate a true charged-particle 
detection. The timing and pulse-height information from the photomultiplier 
tubes at both ends of the sensors are required to be consistent.
TOF hits are then matched to the best extrapolated Spectrometer track. A residual
(minimum distance of TOF hit from extrapolated track trajectory) of less than 
4~cm is required for matching. 
Over a distance of more than 4~m from the last Spectrometer layer to 
the TOF wall, tracks are matched to TOF hits with a resolution of 1.5~cm
(3.5 cm FWHM).

\section{Particle Identification in the Spectrometer}
\label{appendix_sipid}

\subsection{Very low momentum particles}

The determination of particle mass at very low transverse momenta 
(0.03--0.05 GeV/c for pions,
0.09--0.13 GeV/c for kaons and 
0.14--0.21 GeV/c for protons) was based on 
detailed GEANT simulations of the measured energy depositions in the
first detector layers.
The required smallest specific energy loss ($dE/dx$) was
six times that of a minimum ionizing particle (MIP).
Consistency between energy deposited in the different layers,
and consistency between the mean measured $dE/dx$ and the 
expected specific energy loss from the simulations
was used to distinguish between pions, protons and kaons.
Additional cuts on the deviations of the candidate track from a straight
line trajectory (allowing for mass and momentum dependent multiple
scattering) were used to reject fake tracks.
The first five layers of the Spectrometer are located in a magnetic
field smaller than 0.3~T, which is insufficient to cleanly separate
positive and negative particles. Therefore at very low momenta, only the
sum of positive and negative particle yields are presented:
($\pi^++\pi^-$), (K$^+$+K$^-$) and (p+$\overline{\rm p}$).
More details about the low-$p_T$ analysis and corrections applied to
these results can be found in the report on a similar measurement
completed at $\sqrt{s_{_{NN}}}=$200 GeV
\cite{phobos:stopping}.

\subsection{Particles with intermediate momenta}

Identification based on the truncated mean value of the specific energy 
loss ($dE/dx$) in the 
silicon layers was applied to particles with high enough momentum
to safely travel through the whole silicon Spectrometer system.
The total momentum range for this PID method was 0.2--0.9~GeV/c 
for pions and kaons, and 0.3--1.4~GeV/c for 
protons and antiprotons. In the measured momentum range the
$dE/dx \propto 1/v^2$ relation holds, where $v$ is the 
velocity. Thus, particles with the same total momentum lose different amounts of
energy in the spectrometer layers (top panel of Fig.~\ref{pidplot}).


The particles can only be identified individually at low momentum, 
therefore a fit method was applied which significantly 
extends the momentum reach by counting abundances of the various species 
in a statistical sense.
The particles were sorted into total momentum bins and fits to the 
specific energy loss ($dE/dx$) histograms 
(inset in Fig.~\ref{pidplot}) were performed.  
Yields of pions and kaons are quoted only below 0.6 GeV/c
transverse momentum, where they produce sufficiently different 
amount of ionization.
The fits no longer work below 0.2 GeV/c total momentum
because of low statistics caused by low tracking efficiency and acceptance,
and both the efficiency and momentum 
corrections become large and poorly known.

At first, $dE/dx$ (truncated mean of the path length-corrected 
specific ionization values) of each track in a given total momentum ($p$) 
bin are collected. At the same time, the $p$ and transverse momentum 
($p_T$) distributions of these tracks are also collected.

A Gaussian fit to the pion $dE/dx$ peak is performed in the momentum bin 
of $p$ = 0.55$\pm$0.05 GeV/c, where the kaons and pions are well separated, 
to measure the $dE/dx$ resolution. The width is found to be about 0.07 MIP, 
but this overestimates the
resolution because the momentum spread within 
this bin widens the $dE/dx$ distribution.

For the theoretical $dE/dx$ function, a form of the Bethe-Bloch formula
was used where all of the constants in it were collapsed into two 
variables.  One  ($E_0$) is an overall constant, and the other 
($b$=20) characterizes the relative strength of the logarithmic rise:

\begin{equation}
dE/dx = E_0 \beta^{-2} (b+2 \ln \gamma - \beta^2)
\label{bethebloch}
\end{equation}

For each $p$ bin, the convolution of the 
theoretical $dE/dx$ function (between the bin edges) and the 
resolution function gives the line shape for each species.
More precisely, a sum of a Gaussian and a half Gaussian 
both with the mean given by Equation \ref{bethebloch}
but different width and amplitude was used to create a more 
realistic lineshape, to account for the natural remnant of the Landau-tail 
of the $dE/dx$ (truncated mean) distribution. The `main' Gaussian accounts for 
90\% of the total lineshape integral; the other 10\% is accounted for by 
the half Gaussian with twice the width on the positive side of the 
distribution. The $dE/dx$ resolution used is momentum dependent; 
$\sigma(dE/dx) \propto (dE/dx)^{\alpha}$.
The best lineshape fit was found using $\alpha=0.9$. This slightly 
non-Gaussian lineshape  gives a good description of the 
experimental $dE/dx$ histograms (see Fig.~\ref{pidplot}).

After building the lineshapes, the only free parameters are the 
amplitudes (yields) of the three particle species.  
A three-parameter $\chi^2$ fit, using the proper weights (the square-root 
of the bin counts) gives the {\it errors} of the yields correctly, 
including the errors caused by overlaps (correlations between yields).
A second fit, where all the weights are set to unity (independent of 
bin content) gives a more precise estimate of the {\it integral} of the 
line shapes by suppressing the importance of tails 
and outliers of the distribution. Thus, the second fit was used to obtain 
the yields while the first one was used for the statistical error.

This method gives a reliable fit when the particle peaks 
are separate. Above 1.4 GeV/c transverse momentum, 
the pion and kaon peaks merge and 
separate fits to kaons and pions are no longer possible.
However, a lineshape for the meson peak is still needed to be able to  
fit the proton (and meson) yield. Here, a certain $K/\pi$ ratio (as a 
function of $p_T$) had to be assumed to build the meson lineshape.  The 
$K/\pi$ ratio was measured as a function of $p_T$ below 1.4 GeV/c, 
where the particles are still separated, and extrapolated to 
higher $p_T$ values. Any particular extrapolation gets less justified
with increasing momentum, but at the same time, the assumed $K/\pi$ ratio  
affects the line shape description less, since only a very small
difference remains between the mean $dE/dx$ value expected for kaons and pions.
Thus, the systematic error 
on the proton yield caused by the uncertainty of the meson line 
shape depends only weakly on $p_T$, and  was found to be a few 
percent.


\section{Corrections}
\label{appendix_corrections}

The corrections applied to the invariant particle yields
are discussed in the following sections.

\subsection{Geometrical Acceptance and Tracking Efficiency}

The most important corrections to the raw particle yields are to account 
for the geometrical acceptance of the detector and the momentum dependence 
of the track-reconstruction efficiency.
These two corrections are combined in a Monte Carlo study of single
particles passing through a GEANT simulation of the PHOBOS detector and 
the full track reconstruction package. This gives the probability, as a 
function of transverse momentum, that an individual particle will be 
properly reconstructed. These functions, obtained for each particle species
over the whole phase space, are used to correct the raw distributions.

\subsection{Occupancy Correction in the Spectrometer}

The track-reconstruction efficiency is to a small extent dependent on the 
occupancy of the Spectrometer. This effect is studied  by embedding and
reconstructing individual Monte Carlo tracks in real data events. 
For central Au+Au collisions at 62.4~GeV, the difference relative to the 
single track case was found to be $\sim$10\%, independent of momentum.


\subsection{Ghost Correction}

After accounting for the efficiency of the track-reconstruction procedure,
we must also account for its purity, since it is possible that it will produce
spurious tracks, called `ghosts.'
Ghost contributions depend on both the track momentum and the 
hit density and are determined by reconstructing Monte Carlo 
events, where the tracking output can be compared to the known input tracks.
The ghost fraction is found to range from roughly 2\% for the most peripheral 
events used in this analysis to 5\% for the most central, with an approximately
exponential dependence on transverse momentum. No species dependence was 
found.

\subsection{Momentum Resolution Correction}

The momentum resolution of the Spectrometer in our momentum range (up to 
$p_T=3$ GeV/c) is a few percent. The correction of the 
invariant yields for this momentum resolution 
depends on the steepness of the momentum distributions, and can be 
estimated to be 2-3\% at $p_T=3$ GeV/c, where it reaches its maximum.
However, instead of an explicit momentum resolution correction,
an implicit one was used in the present analysis.
The efficiency of the detector system was estimated using Monte Carlo 
techniques, where single tracks weighted by the closely exponential 
transverse momentum distribution were generated, and reconstructed (with 
the momentum resolution entering the procedure at this step). 
The reconstructed MC track sample was analyzed the same way as the data
and compared to the originally assumed exponential $p_T$ distribution. 
Therefore the momentum resolution correction is taken care of by the 
efficiency correction in an implicit way.

\subsection{Feed-down Correction}

Contributions to the measured transvserse momentum spectra from particles which are products of weakly-decaying resonance states are detector-dependent, and therefore should be removed.

\subsubsection{Proton Feed-down}

For a reconstructed proton we need to know the
probability that it is a primary particle and not the product of a weak
decay.
To obtain this, one needs to know the relative efficiency of 
reconstructing primary protons versus those which are products of weak 
decays, and also know the yield of primary protons relative to particles 
which can produce protons by weak decay.


The major feed-down contribution to the proton yields comes from the $\Lambda$ decay: $\Lambda \rightarrow p + \pi^-$. This process has a branching ratio of 63.9\% and a lifetime expressed as $c\tau = 7.9$~cm; the daughter particles have a momentum of 101~MeV/c in 
the center of mass frame.

The $\Lambda$, being neutral, leaves no trace as it passes through the 
silicon detectors. The PHOBOS tracking procedures are such that the 
daughter proton will only be reconstructed if the decay happens before 
the first spectrometer layer. As this layer is only roughly 10~cm 
from the nominal interaction point, the PHOBOS experiment has 
sensitivity to distinguish between primary and feed-down protons.

The GEANT Monte Carlo package was used to simulate single $\Lambda$ decays 
in the PHOBOS spectrometer. 
$\Lambda$s were generated with realistic transverse momentum 
distributions, and the efficiency of reconstructing the daughter proton 
at a given momentum can then be compared to the efficiency for primary 
proton reconstruction at the same momentum. This result can then be 
scaled according to the $\Lambda/p$ ratio to determine, as a function of 
transverse momentum, the fraction of observed protons which are expected 
to actually originate from $\Lambda$ decays.
This method was tested on Monte Carlo events produced by the HIJING event 
generator. 
In this case, the $\Lambda/p$ ratio was known, and this method was able to 
correctly describe the contribution from feed-down protons.

The PHENIX collaboration has measured $\Lambda/p = 0.89 \pm 0.07$ in 
Au+Au collisions at $\sqrt{s_{_{NN}}} = 130$~GeV \cite{phenix:lambda}.
The value at 62.4~GeV is not yet known, which leads to an uncertainty in 
the feed-down correction. 
Studies were performed for a variety of different values of the ratio in 
the range $0.7 \le \Lambda/p \le 1.1$.

Feed-down protons can also originate from the decay $\Sigma^+ \rightarrow 
p + \pi^0$, which has a branching ratio of 51.6\% and $c\tau = 2.4$~cm.
The $\Sigma/p$ ratio has not been measured for Au+Au collisions at RHIC. 
The HIJING event generator predicts $\Sigma/p \approx 0.3$, and 
measurements from p+p collisions at similar energies have found a value 
of around 0.5.
Monte Carlo studies of single $\Sigma$ decays in the PHOBOS 
detector were performed using the same 
techniques as for $\Lambda$s, taking the ratio $\Sigma/p$ in the range 
$0.1 \le \Sigma/p \le 0.9$.
For the final correction the $\Lambda/p = 0.9$ and $\Sigma/p = 0.3$ 
ratios were used, as discussed in the following paragraphs.


The distance of closest approach of a reconstructed track to the event 
vertex (DCA) is a quantity which has sensitivity for distinguishing 
between primary particles and those which did not originate from the event 
vertex. This is particularly useful for the study of feed-down 
particles.
From the Monte Carlo simulations described above, the DCA distributions for 
primary and feed-down protons were obtained. 
In the case of primaries, the DCA distribution is narrow, reflecting the fact 
 that these particles really did originate at the event vertex. The 
 distribution for feed-down particles has a tail which extends to much higher 
 values of DCA.
 The cut on DCA $< 0.35$~cm used as part of track selection was found to 
 remove $\sim 25$\% of the feed-down protons but only $\ll 1$\% of the 
 primaries.

\begin{figure}[t]
\centerline{
\includegraphics[width=43mm]{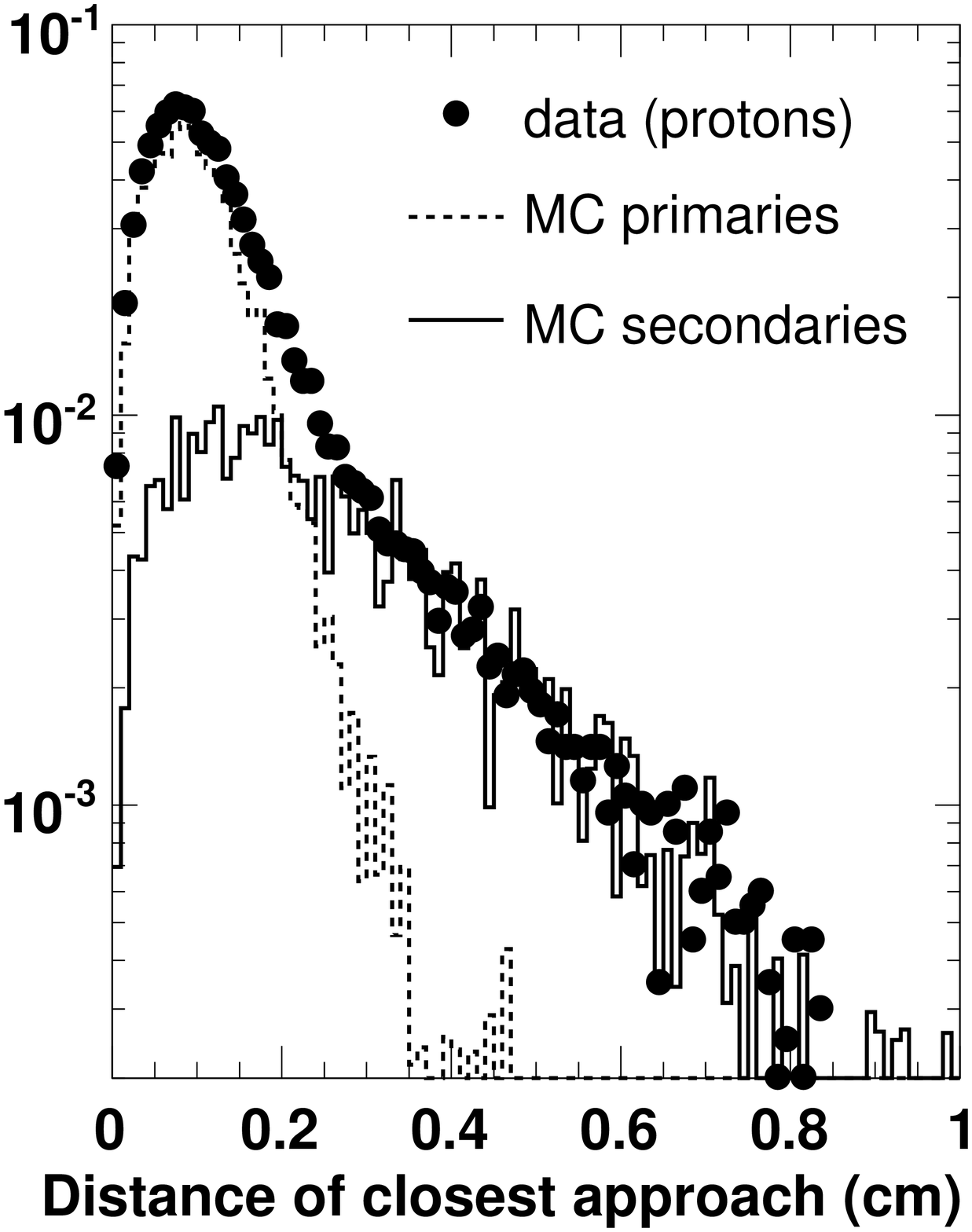}
\includegraphics[width=43mm]{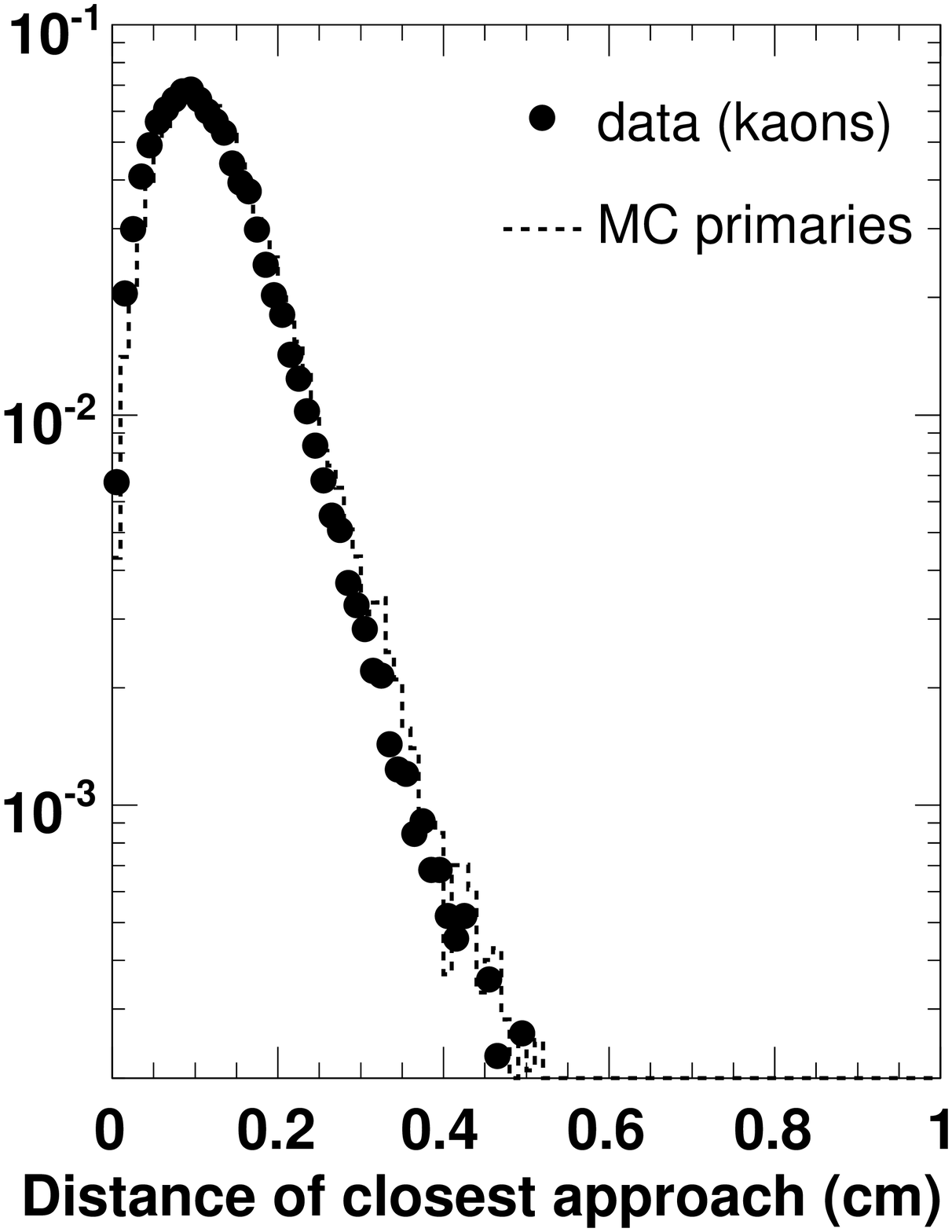}
}
\caption{\label{dca_study}
Distribution of the distance of closest approach between the track and the 
event vertex for protons (left panel) and for kaons (right panel). The 
dots are identified protons and kaons from the data,
the histograms are Monte Carlo-simulated primary protons and kaons (dashed 
lines) and secondaries (weak decay daughters, solid lines) assuming 25\% 
secondary contribution below 0.35~cm for the protons.}
\end{figure}

The simulated primary and feed-down DCA distributions were combined to 
reproduce the observed distribution for data protons. It was found that a 
feed-down contribution in the range of 25-30\% gave the best consistency, 
and that less than 20\% or greater than 35\% seemed to be inconsistent with 
the data. The left panel of Figure \ref{dca_study} illustrates the proton 
DCA distribution for the identified protons in the data (dots), compared 
to simulated primary and weak decay daughter (secondary) protons, where 
the contribution of secondaries is set to 25\%. On the right panel,
the identified kaon DCA distribution is plotted, together with the 
simulated primary kaon DCA distribution. The agreement confirms that
there are no secondary kaons originating from 
weak decays visible in the data.

 The final feed-down correction for protons is chosen to be the sum of the 
 simulated contributions from $\Lambda$ and $\Sigma$ decays, assuming 
 $\Lambda/p = 0.9$ and $\Sigma/p = 0.3$. This gives a correction which is 
 consistent with the data-driven feed-down estimates
 from the DCA distributions. The systematic uncertainty on 
 this correction comes from plausible variations in the $\Lambda/p$ and 
 $\Sigma/p$ ratios (which have not yet been measured for $62.4$~GeV Au+Au 
 collisions) and are also consistent with the bounds obtained from the DCA 
 analysis.

 The function used for the fraction of observed protons which are feed-down 
 products is:
\begin{equation}
f(p_T) = 0.39e^{-0.57 p_T}+(0.22p_T-0.16)e^{-5p_T^2}
\end{equation}
 with upper and lower bounds defined by $(1.22+0.1p_T)f(p_T)$ and 
 $(0.78-0.1p_T)f(p_T)$ respectively.

\subsubsection{Antiproton Feed-down}

We determine the antiproton feed-down correction in relation to that for protons by making the reasonable assumption
that antiproton feed-down is dominated by $\bar{\Lambda}$ decays
and therefore the main difference between proton and antiproton feed-down comes from differences in the $\bar{\Lambda}/\bar{p}$ and $\Lambda/p$ ratios.

The value of these ratios in Au+Au collisions at $\sqrt{s_{_{NN}}} = $130~GeV has been measured \cite{phenix:lambda} to be:
\[
\frac{\Lambda}{p} = 0.89 \pm 0.07 \mbox{(stat)}
\]
\[
\frac{\bar{\Lambda}}{\bar{p}} = 0.95 \pm 0.09 \mbox{(stat)}
\]

By considering the quark content of these states, one can postulate that they should be related by:
\begin{equation}
\frac{\bar{\Lambda}}{\bar{p}} = \frac{\Lambda}{p} \times \frac{K^+}{K^-}
\end{equation}
This relationship was found to hold true for 130 GeV Au+Au collisions, where the kaon ratio was measured \cite{phobos:pbarp_AuAu_130} to be:
$K^-/K^+ = 0.91 \pm 0.07 \mbox{(stat)} \pm 0.06 \mbox{(sys)}$.

We assume it holds at 62.4 GeV also, where we have made preliminary 
measurements of $K^-/K^+ \approx 0.85$. Thus we take the antiproton 
feed-down correction to be roughly $1.0/0.85 = 1.18$ times the correction 
for protons.

\subsubsection{Feed-down to Kaons and Pions}

There are no weak processes which produce kaons as the final state,
thus there is no feed-down correction for the kaon yields. 
This expectation from theoretical grounds is verified by the observed DCA distribution for reconstructed kaons. 
The data agree well with the distribution for primary kaons obtained from Monte Carlo simulations, with no indication of a tail that would correspond to feed-down contributions.

The DCA distribution for pions from data was also studied and compared to that for simulated primary pions. 
The feed-down contribution to the pion yields after applying the cut of DCA $< 0.35$~cm was estimated to be less than 1\% and was therefore neglected, but this was included as a 1\% contribution to the overall systematic uncertainty on the pion spectra.

\subsection{Secondaries}

Secondary particles are here defined as those which did not originate directly
from the heavy-ion collision and are not the product of weak decays. 
The main sources of secondaries are interactions of primary particles with 
the beryllium beampipe and other detector material.

Production of secondary particles is studied using a GEANT simulation of the 
PHOBOS detector, with the HIJING event generator as the source of primary
particles. 
On average, secondary particles tend to be produced with very low momentum. 
This means that they are unlikely to be detected in our Spectrometer because 
they tend to bend out of the acceptance and also suffer more multiple 
scatterings, making their hit pattern less likely to be reconstructed
as a track.

Secondaries were only found to contribute significantly to the total proton 
yield for $p_T < 200$~MeV/c. 
As this analysis does not extend this low in transverse momentum,
the secondary correction to our measured proton yield is actually 
less than 1\%.
We therefore choose to neglect this correction, but make an additional 
contribution of 1\% to the systematic uncertainty of the spectra.
The secondary contribution to kaon and pion yields was found to be 
negligible.

\subsection{Dead Channels in the Spectrometer and in the TOF}

Some of the electronics channels in the silicon Spectrometer
do not produce a signal when a charged particle passes through the
silicon pad attached to it, and some other channels are called `hot' 
since they produce a signal even without particles crossing.
The fraction of these faulty channels is on the percent level.
They are identified by inspecting the energy deposit ($dE$) distributions
channel by channel, where the above failures of operation are found.
The set of these faulty channels are excluded from the analysis.

Technically, the full data sample is reconstructed by using all the 
channels, and a separate dead channel correction is applied.
The correction is obtained by reconstructing part of the data, as well as 
the Monte Carlo track sample again with the dead/hot channels excluded. 
Since the exclusion of channels decreases the number of found tracks 
in both the data and the Monte Carlo samples (where the latter is used to 
quantify the geometrical acceptance and efficiency), the dead channel 
correction is applied as the ratio of the change of the track yields in 
the two cases.

Another type of dead channel correction is performed on the number of 
tracks detected by the Time of Flight wall for the part of data that uses 
the TOF information. Four out of the 120 channels of the TOF wall are
dead, and there are no hot channels. By comparing the hit frequency 
distribution of the 116 live channels (which is a linear function of the 
channel number to a good approximation) to the position of the missing 
channels, the reduction of measured yield for the entire wall caused 
by the missing channels is estimated to be 3\%, and that correction 
is applied to the data.



\end{document}